\documentclass{lmcs} 
\pdfoutput=1

\usepackage{lastpage}
\lmcsdoi{18}{1}{9}
\lmcsheading{}{\pageref{LastPage}}{}{}%
{Dec.~25,~2020}{Jan.~12,~2022}{}

\usepackage[utf8]{inputenc}


\usepackage{stmaryrd}
\usepackage{proof}
\usepackage{amssymb}
\usepackage{mathpartir}
\usepackage{ntabbing}
\usepackage{listings}
\usepackage{booktabs}
\usepackage{bookmark}

\newlength{\rWidth}

\newcommand{\m}[1]{\mathsf{#1}}
\newcommand{\mb}[1]{\mathbf{#1}}

\newenvironment{sill}{\begin{quote}\begin{tabbing}}{\end{tabbing}\end{quote}}

\newcommand{\Sg}{\Sigma}
\newcommand{\xvdash}[1]{%
  \vdash^{\mkern-8mu\scriptstyle\rule[-1.2ex]{0pt}{0pt}#1}%
}

\newcommand{\D}{\Delta}

\newcommand{\proves}{\vDash}

\newcommand{\proc}[2]{\m{proc}(#1, #2)}
\newcommand{\msg}[2]{\m{msg}(#1, #2)}

\newcommand{\fresh}[1]{(#1 \text{ fresh})}

\newcommand{\ecase}[3]{\m{case} \; #1 \; (#2 \Rightarrow #3)}

\newcommand{\erecvch}[2]{#2 \leftarrow \m{recv} \; #1}
\newcommand{\erecvn}[2]{\{#2\} \leftarrow \m{recv} \; #1}

\newcommand{\esendch}[2]{\m{send} \; #1 \; #2}
\newcommand{\esendn}[2]{\m{send} \; #1 \; \{#2\}}

\newcommand{\ewait}[1]{\m{wait} \; #1}

\newcommand{\eclose}[1]{\m{close} \; #1}

\newcommand{\fwd}[2]{#1 \leftrightarrow #2}

\newcommand{\esendl}[2]{#1.#2}

\newcommand{\ecut}[4]{#1 \leftarrow #2 \;\; #3 \semi #4}

\newcommand{\edelay}[1]{\m{delay} \, (#1)}
\newcommand{\ewhen}[1]{\m{when?} \, (#1)}
\newcommand{\enow}[1]{\m{now!} \, (#1)}

\newcommand{\noww}{\m{now!}}
\newcommand{\whenn}{\m{when?}}
\newcommand{\ework}[1]{\m{work} \, \{#1\}}
\newcommand{\eget}[2]{\m{get} \, #1 \, \{#2\}}
\newcommand{\epay}[2]{\m{pay} \, #1 \, \{#2\}}
\newcommand{\procdef}[3]{#3 \leftarrow #1 \; #2}

\newcommand{\eassume}[2]{\m{assume} \; #1 \; \{#2\}}
\newcommand{\eassert}[2]{\m{assert} \; #1 \; \{#2\}}

\newcommand{\eimposs}{\m{impossible}}

\newcommand{\lolli}{\multimap}
\newcommand{\tensor}{\otimes}
\newcommand{\with}{\mathbin{\binampersand}}

\newcommand{\one}{\mathbf{1}}

\newcommand{\semi}{\; ; \;}
\newcommand{\ichoiceop}{\oplus}
\newcommand{\echoiceop}{\with}
\newcommand{\ichoice}[1]{\ichoiceop \{ #1 \}}
\newcommand{\echoice}[1]{\echoiceop \{ #1 \}}

\newcommand{\mi}[1]{\mbox{\it #1}}

\newcommand{\tassertop}{{?}}  
\newcommand{\tassumeop}{{!}}  
\newcommand{\tassert}[1]{\tassertop\{#1\}. \,} 
\newcommand{\tassume}[1]{\tassumeop\{#1\}. \,} 

\newcommand{\entailpot}[1]{\xvdash{#1}}

\newcommand{\paypot}{\triangleright}
\newcommand{\getpot}{\triangleleft}
\newcommand{\tgetpot}[2]{\getpot^{#2} #1}
\newcommand{\tpaypot}[2]{\paypot^{#2} #1}

\newcommand{\tdia}[1]{\Diamond #1}
\newcommand{\tbox}[1]{\Box #1}
\newcommand{\Dia}{\Diamond}
\newcommand{\tforall}[1]{\forall #1. \, }
\newcommand{\texists}[1]{\exists #1. \, }

\newcommand{\Next}{\raisebox{0.3ex}{$\scriptstyle\bigcirc$}}
\renewcommand{\next}[1]{\Next #1}
\newcommand{\tdelay}[2]{
    \IfEqCase{#2}{%
        {1}{\next{#1}}%
    }[{\Next^{#2} (#1)}]%
}%

\newcommand{\lforce}[2]{[#1]_L^{#2}}
\newcommand{\rforce}[2]{[#1]_R^{#2}}

\newcommand{\indv}[1]{[\overline{#1}]}

\newcommand{\indn}[1]{\{\overline{#1}\}}

\newcommand{\config}{\mathcal{S}}

\newcommand{\delayedbox}[1]{#1 \; \m{delayed}^{\Box}}
\newcommand{\delayeddia}[1]{#1 \; \m{delayed}^{\Diamond}}

\newcommand{\valid}[1]{#1 \; \m{valid}}

\newcommand{\cons}{\mathcal{C}}
\newcommand{\vars}{\mathfrak{v}}
\newcommand{\Vars}{\mathcal{V}}

\definecolor{verylightgray}{rgb}{.97,.97,.97}

\lstdefinelanguage{Solidity}{
	keywords=[1]{anonymous, assembly, assert, balance, break, call, callcode, case, catch, class, constant, continue, constructor, contract, debugger, default, delegatecall, delete, do, else, emit, event, experimental, export, external, false, finally, for, function, gas, if, implements, import, in, indexed, instanceof, interface, internal, is, length, library, log0, log1, log2, log3, log4, memory, modifier, new, payable, pragma, private, protected, public, pure, push, require, return, returns, revert, selfdestruct, send, solidity, storage, struct, suicide, super, switch, then, this, throw, transfer, true, try, typeof, using, value, view, while, with, addmod, ecrecover, keccak256, mulmod, ripemd160, sha256, sha3}, 
	keywordstyle=[1]\color{blue}\bfseries,
	keywords=[2]{address, bool, byte, bytes, bytes1, bytes2, bytes3, bytes4, bytes5, bytes6, bytes7, bytes8, bytes9, bytes10, bytes11, bytes12, bytes13, bytes14, bytes15, bytes16, bytes17, bytes18, bytes19, bytes20, bytes21, bytes22, bytes23, bytes24, bytes25, bytes26, bytes27, bytes28, bytes29, bytes30, bytes31, bytes32, enum, int, int8, int16, int24, int32, int40, int48, int56, int64, int72, int80, int88, int96, int104, int112, int120, int128, int136, int144, int152, int160, int168, int176, int184, int192, int200, int208, int216, int224, int232, int240, int248, int256, mapping, string, uint, uint8, uint16, uint24, uint32, uint40, uint48, uint56, uint64, uint72, uint80, uint88, uint96, uint104, uint112, uint120, uint128, uint136, uint144, uint152, uint160, uint168, uint176, uint184, uint192, uint200, uint208, uint216, uint224, uint232, uint240, uint248, uint256, var, void, ether, finney, szabo, wei, days, hours, minutes, seconds, weeks, years},	
	keywordstyle=[2]\color{teal}\bfseries,
	keywords=[3]{block, blockhash, coinbase, difficulty, gaslimit, number, timestamp, msg, data, gas, sender, sig, value, now, tx, gasprice, origin},	
	keywordstyle=[3]\color{violet}\bfseries,
	identifierstyle=\color{black},
	sensitive=false,
	comment=[l]{//},
	morecomment=[s]{/*}{*/},
	commentstyle=\color{gray}\ttfamily,
	stringstyle=\color{red}\ttfamily,
	morestring=[b]',
	morestring=[b]" 
}

\theoremstyle{plain} 

\begin{document}

\title[Rast]{Rast: A Language for Resource-Aware Session Types}
\titlecomment{short version accepted to FSCD 2020.}

\author[A.~Das]{Ankush Das\rsuper{a,*}}	
\address{Amazon, Cupertino, CA, USA}
\email{ankushd@cs.cmu.edu}
\thanks{\lsuper{*}Research conducted prior to joining Amazon}

\author[F.~Pfenning]{Frank Pfenning\rsuper{b}}	
\address{Carnegie Mellon University}	
\email{fp@cs.cmu.edu}  
\thanks{supported by the National
Science Foundation under SaTC Award 1801369,
CAREER Award 1845514, and Grant No. 1718276.}	




\begin{abstract}
  Traditional session types prescribe bidirectional communication
  protocols for concurrent computations, where well-typed programs are
  guaranteed to adhere to the protocols.
  However, simple session types cannot capture properties beyond the
  basic type of the exchanged messages.
  In response, recent work has extended session types with refinements
  from linear arithmetic, capturing intrinsic attributes of processes and data.
  These refinements then
  play a central role in describing sequential and parallel complexity
  bounds on session-typed programs.
  The Rast language provides an open-source implementation
  of session-typed concurrent programs extended with arithmetic
  refinements as well as ergometric and temporal types to capture work
  and span of program execution.
  To further support generic programming, Rast also enhances arithmetically
  refined session types with recently developed nested parametric polymorphism.
  Type checking relies on Cooper's
  algorithm for quantifier elimination in Presburger arithmetic with a
  few significant optimizations, and a heuristic extension to
  nonlinear constraints.  Rast furthermore includes a reconstruction
  engine so that most program constructs pertaining the layers of
  refinements and resources are inserted automatically.  We provide
  a variety of examples to demonstrate the expressivity of the
  language.
\end{abstract}

\maketitle

\section{Introduction}\label{sec:intro}

\emph{Session types}~\cite{Honda93concur,Honda98esop,Vasconcelos12ST}
provide a structured way of statically prescribing communication
protocols in message-passing programs. In this paper, we
introduce the Rast programming language and implementation which is based
on \emph{binary session types} governing the interaction of two
processes along a single channel, rather than \emph{multiparty
session types}~\cite{Honda08popl} which take a more global view of
computation.  Nevertheless, during the execution of a Rast program,
complex networks of interacting processes arise.  Recent work has
placed binary session types without general recursion on a strong
logical foundation by exhibiting a Curry-Howard isomorphism with
linear logic~\cite{Caires10concur,Wadler12icfp,Caires16mscs}.
Due to this correspondence, the cut reduction properties of linear logic entail type
safety of session typed processes and guarantee \emph{freedom from
deadlocks} (global progress) and \emph{session fidelity} (type
preservation) ensuring adherence to the communication protocols at
runtime.

The Rast programming language is based
on session types derived from intuitionistic linear logic, extended
with equirecursive types and recursive process definitions.
Rast also supports full parametric polymorphism
enabling definition of polymorphic data structures. It
furthermore supports arithmetic type refinements as well as ergometric
and temporal types to measure the total work and span of Rast
programs.
The theory underlying Rast
has been developed in several papers, including the Curry-Howard
interpretation of linear logic as session-typed
processes~\cite{Caires10concur,Caires16mscs}, the treatment of
general equirecursive types and type equality~\cite{Gay2005},
asynchronous communication~\cite{Gay10jfp,DeYoung12csl}, ergometric
types~\cite{Das18lics}, temporal types~\cite{Das18icfp}, indexed
types~\cite{Griffith13nfm,Das20PPDP}, indexed type equality~\cite{Das20CONCUR},
and nested polymorphism~\cite{Das21ESOP}.


We focus on key aspects of language design and implementation, not the
underlying theory which can be found in the cited papers.  A notable
exception is subtyping for the full language, including nested
polymorphic types, whose properties are the subject of ongoing
research.  We present Rast in layers, using typing rules, an
operational semantics, and examples to explain and illustrate
increasingly advanced features via their type structure.  All language
layers satisfy the properties of preservation (session fidelity) and
progress (deadlock-freedom), with slightly different statements
depending on the semantic properties under consideration.  For
example, in the presence of ergometric types the sum of potential and
total work expended remains constant.  This paper is a significantly
revised and extended version of a system description~\cite{Das20FSCD},
including the formal definition of typing and computation, nested
parametric polymorphism, subtyping, and additional examples.

We begin with motivation and a brief overview of the main features of
the language using a concurrent queue data structure as a running
example.  The following type specifies the interface to a polymorphic queue server
in the system of basic recursive session types storing elements of type $A$
and supporting the operations of insert and delete.

\begin{sill}
$\m{queue}[A] = \echoice{$\=$\mb{ins} : A \lolli \m{queue}[A],$ \\
\>$\mb{del} : \ichoice{$\=$\mb{none} : \one,$\\
\>\>$\mb{some} : A \tensor \m{queue}[A]}}$
\end{sill}

The \emph{external choice} operator $\echoiceop$ dictates
that the process providing this data structure accepts either one of
two messages: the labels $\mb{ins}$ or $\mb{del}$.  In the case of
$\mb{ins}$, it receives an element of type $A$ denoted by the $\lolli$
operator, and then the type recurses back to $\m{queue}[A]$. On receiving
a $\mb{del}$ request, the process can respond with one of two labels
($\mb{none}$ or $\mb{some}$), indicated by the \emph{internal choice}
operator $\ichoiceop$.  If the queue is empty, it responds with
$\mb{none}$ and then \emph{terminates} (indicated by $\one$).  If the
queue is nonempty, it responds with $\mb{some}$ followed by the
element of type $A$ (expressed with the $\tensor$ operator) and
recurses.
The type $\m{queue}[A]$ denotes the type name $\m{queue}$ instantiated
with the session type $A$.

However, the simple session type does not express the
conditions under which the $\mb{none}$ and $\mb{some}$ branches must
be chosen, which requires tracking the length of the queue.
Rast extends session types with arithmetic
refinements~\cite{Das20CONCUR,Das20PPDP} which can be used to express the length of a
queue. The more precise type
\begin{sill}
$\m{queue}[A]\{n\} = \echoice{$\=$\mb{ins} : A \lolli \m{queue}[A]\{n+1\},$\\
\>$\mb{del} : \ichoice{$\=$\mb{none} : \tassert{n=0} \one,$\\
\>\>$\mb{some} : \tassert{n > 0} A \tensor \m{queue}[A]\{n-1\}}}$
\end{sill}
uses the index refinement $n$ to indicate the number of elements in
the queue.
In addition, the \emph{type constraint} $\tassert{\phi} A$
read as ``\textit{there exists a proof of $\phi$}'' is analogous to
the \emph{assertion} of $\phi$ in imperative languages. Conceptually,
the process providing the queue must provide a proof of $n = 0$ after
sending $\mb{none}$, and a proof of $n > 0$ after sending $\mb{some}$
respectively.  It is therefore constrained in its choice between the
two branches based on the value of the index $n$.  Since the constraint
domain is decidable and the actual form of a proof is irrelevant to
the outcome of a computation, in the implementation no proof is actually
sent.

As is standard in session types, the dual constraint to
$\tassert{\phi} A$ is $\tassume{\phi} A$ (\textit{for all proofs of
$\phi$}, analogous to the \emph{assumption} of $\phi$). We also add
explicit quantifier type constructors $\texists{n} A$ and $\tforall{n} A$
that send and receive natural numbers, respectively.

Arithmetic refinements are instrumental in expressing
\emph{sequential} and \emph{parallel complexity bounds}.  These are
captured with ergometric~\cite{Das18lics,Das21CSF} and temporal
session types~\cite{Das18icfp}.  They rely on index refinements to
express, for example, the size of lists, stacks, and queue data
structures, or the height of trees and express work and time bounds as
a function of these refinements.


Ergometric session types~\cite{Das18lics} capture the sequential
complexity of programs, often called the \emph{work}.
Revisiting the queue example, consider an implementation where each
element in the queue corresponds to a process.  Then insertion acts
like a bucket brigade, passing the new element one by one to the end
of the queue.  Among multiple cost models provided by Rast is one
where each \emph{send} operation requires $1$ unit of work (erg).  In
this cost model, such a bucket brigade requires $2n$ ergs because each
process has to send $\mb{ins}$ and then the new element.
On the other hand, responding to the $\mb{del}$ request requires
only $2$ ergs: the provider responds with $\mb{none}$ and closes the channel,
or $\mb{some}$ followed by the element.  This gives us the following
type
\begin{sill}
  $\m{queue}[A]\{n\} = \echoice{$\=$\mb{ins} : \textcolor{red}{\getpot^{2n}}
  (A \lolli \m{queue}[A]\{n+1\}),$\\
  \>$\mb{del} : \textcolor{red}{\getpot^{2}} \ichoice{$\=$\mb{none} :
  \tassert{n=0} \one,$\\
  \>\>$\mb{some} : \tassert{n > 0} A \tensor \m{queue}[A]\{n-1\}}}$
\end{sill}
which expresses that the client has to send $2n$ ergs of potential to insert an
element ($\textcolor{red}{\getpot^{2n}}$), and $2$ ergs to delete an
element ($\textcolor{red}{\getpot^{2}}$).
The ergometric type system (described in Section~\ref{sec:ergo})
verifies this work bound using the potential operators as described in
the type.

Temporal session types~\cite{Das18icfp} capture the time
complexity of programs assuming maximal parallelism on
unboundedly many processors, often called the \emph{span}.  How does
this work out in our example?  We adopt a cost model where each send
and receive action takes one unit of time (tick).  First, we note that
a use of a queue is at the client's discretion, so should be available
at \emph{any} point in the future, expressed by the type constructor
$\Box$.  Secondly, the queue does not interact at all with the
elements it contains, so they have to be of type $\Box A$ for an
arbitrary $A$.  Since each interaction takes 1 tick, the next
interaction requires at least 1 tick to elapse, captured by the
next-time operator $\Next$.  During insertion, we need more time than this:
a process needs $2$ ticks to pass the element down the queue, so it
takes $3$ ticks overall until it can receive the next insert or delete
request after an insertion.  This reasoning yields the following
temporal type:

\begin{sill}
  $\m{queue}[A]\{n\} = \textcolor{red}{\Box} \echoice{$\=$\mb{ins} : \textcolor{red}{\Next}
  (\textcolor{red}{\Box}A \lolli \textcolor{red}{\next^{3}} \m{queue}[A]\{n+1\}),$\\
  \>$\mb{del} : \textcolor{red}{\Next} \ichoice{$\=$\mb{none} :
    \textcolor{red}{\Next}\; \tassert{n=0} \one,$\\
  \>\>$\mb{some} : \textcolor{red}{\Next}\; \tassert{n > 0} \textcolor{red}{\Box}A \tensor \textcolor{red}{\Next}\, \m{queue}[A]\{n-1\}}}$
\end{sill}

We see that even though the bucket brigade requires much work for
every insertion (linear in the length of the queue), it has a lot of
parallelism because there are only a constant number of required
delays between consecutive insertions or deletions.

Rast follows the design principle that bases an \emph{explicit language}
directly on the correspondence with the sequent calculus for the
underlying logic (such as linear logic, or temporal or ergometric
linear logic), extended with recursively defined types and processes.
Programming in this fully explicit form tends to be unnecessarily
verbose, so Rast also provides an \emph{implicit language} in which
most constructs related to index refinements and amortized work
analysis are omitted.  Explicit programs are then recovered by a
proof-theoretically motivated algorithm~\cite{Das20PPDP} for \emph{reconstruction}
which is sound and complete on valid implicit
programs.

Rast is implemented in SML and available as an open-source repository~\cite{RastBitBucket}.
It allows the user to choose explicit or
implicit syntax and the exact cost models for work and time analysis.
The implementation consists of a lexer, parser, reconstruction
engine, arithmetic solver, type checker,
and an interpreter, with particular attention
to providing precise error messages.
The repository also contains a number of illustrative
examples that highlight various language features, some of which we
briefly sketch in this paper.

To summarize, Rast makes the following contributions:
\begin{enumerate}
\item A session-typed programming language with arithmetic refinements
  applied to ergometric and temporal types for parallel complexity
  analysis.
\item An extension with full parametric polymorphism to enable generic programming.
\item A subtyping algorithm that works well in practice despite
  its theoretical undecidability~\cite{Das20CONCUR} and uses
  Cooper's algorithm~\cite{cooper1972theorem} with some small
  improvements to decide constraints in Presburger arithmetic (and
  heuristics for nonlinear constraints).
\item A type checking algorithm that is sound and complete relative to
  subtyping.
\item A sound and complete reconstruction algorithm for a process
  language where most index and ergometric constructs remain implicit.
\item An interpreter for executing session-typed programs using the
  recently proposed shared memory semantics~\cite{Pruiksma20arxiv}.
\end{enumerate}


\section{Example: An Implementation of Queues}

\begin{figure}[t]
\begin{lstlisting}[caption={Declaration and definition of queue processes, file \texttt{examples/list.rast}},label=queue_impl,captionpos=b]
type queue[A]{n} = &{ins : A -o queue[A]{n+1},
                       del : +{none : ?{n = 0}. 1,
                                some : ?{n > 0}. A * queue[A]{n-1}}}

decl empty[A] : . |- (q : queue[A]{0})
decl elem[A]{n} : (x : A) (t : queue[A]{n}) |- (q : queue{n+1})

proc q <- empty[A] =
  case q (                              % receive a label along q
   ins => x <- recv q ;                 % if `ins', receive a channel x along q
           e <- empty[A] ;              % spawn a new empty process (at type A)
           q <- elem[A]{0} x e          % continue as an elem holding x
 | del => q.none ;                      % if `del', respond with label `none'
           assert q {0=0} ;             % assert that (n = 0)[0/n]
           close q )                    % terminate by closing q

proc q <- elem[A]{n} x t =
  case q (                              % receive a label along q
   ins => y <- recv q ;                 % if `ins', receive a channel y along q
           t.ins ;                      % send label 'ins' along t
           send t y ;                   % send the channel y along t
           q <- elem[A]{n+1} x t        % continue as elem at index n+1
 | del => q.some ;                      % if `del', respond with label `some'
           assert q {n+1>0} ;           % assert that (n > 0)[n+1/n]
           send q x ;                   % send x along q
           q <-> t )                    % identify q with t and terminate
\end{lstlisting}
\end{figure}

We use the implementation of queues as sketched in the introduction as
a first example program, starting with the indexed version. The
concrete syntax of types is a straightforward rendering of their
abstract syntax (Table~\ref{tab:syntax}).
\begin{lstlisting}
  type queue[A]{n} = &{ins : A -o queue[A]{n+1},
                         del : +{none : ?{n = 0}. 1,
                                  some : ?{n > 0}. A * queue[A]{n-1}}}
\end{lstlisting}
Each channel has exactly two endpoints: a \emph{provider} and a
\emph{client}.  Session fidelity ensures that provider and client
always agree on the type of the channel and carry out complementary
actions. The type of the channel evolves during communication, since
it has to track where the processes are in the protocol as they
exchange messages.

In our example, we need two kinds of processes: an \emph{empty}
process at the end of the queue, and an \emph{elem} process that holds
an element $x$.  The empty process \emph{provides} an empty queue,
that is, a service of type \verb'queue[A]{0}' along a channel named
\verb'q'.  It does not \emph{use} any channels
(indicated by an empty context `\verb'.''), so its type is declared with
\begin{lstlisting}
  decl empty[A] : . |- (q : queue[A]{0})
\end{lstlisting}
Polymorphism is explicit in Rast, requiring the programmer to specify
the free type variables. The declaration \verb'empty[A]' denotes that
the type \verb'A' may occur freely in its type and definition.

An \emph{elem} process \emph{provides} a service of type
\verb'queue[A]{n+1}' along a channel named \verb'q' and \emph{uses} a
queue of type \verb'queue[A]{n}' along a channel named \verb't'.  In
addition, it holds (``owns'') an element \verb'x' of type \verb'A'.
\begin{lstlisting}
  decl elem[A]{n} : (x : A) (t : queue[A]{n}) |- (q : queue[A]{n+1})
\end{lstlisting}
The turnstile `\verb'|-'' separates the channels used from the channel
that is provided (which is always exactly one, analogous to a
value returned by a function).  The notation \verb'elem[A]{n}' indicates
that the type \verb'A' and natural number \verb'n' are parameters
of this process.

Listing~\ref{queue_impl} shows the implementation of the two forms of
processes in Rast.  Comments, starting with a \verb'%' character and
extending to the end of the line, provide a brief explanation for the
actions of each line of code.  This code is in \emph{explicit} form,
both in its refinements and polymorphism.
Thus, the type and natural number parameters to a process need to
be specified explicitly (e.g., \verb'elem[A]{0}' or \verb'elem[A]{n+1}').
Moreover, the code contains two instances of \verb'assert' to match the constraints
\verb'?{n = 0}' and \verb'?{n > 0}' in the two possible responses to a
delete request.
Rast also provides the programmer with an option to write code in \emph{implicit} form
where the two asserts would be omitted since
they can be read off the type at the corresponding place in the
protocol.  Of course, the type checker verifies that the assertion
is justified and fails with an error message if it is not, whether
the construct is explicit or implicit.
However, even in the implicit form, parameters (both type
and natural numbers) to a process need to be provided to spawn a new
instance of it.
The full implementation can be found in \verb'tests/queues-quant.rast'
in the repository.

\paragraph{\textbf{\textit{Concealing Queue Size}}}
As a final pair of illustrations, we describe how to wrap a sized
queue in an unsized one, and vice-versa.
First, we define the type of an unsized queue (introduced in Section~\ref{sec:intro}).
\begin{lstlisting}
  type uqueue[A] = &{ins : A -o uqueue[A],
                       del : +{none : 1,
                                some : A * uqueue[A]}}
\end{lstlisting}
Next, we define a \verb'conceal' process that uses a sized queue
and provides an unsized queue, thus hiding, but \emph{not discarding}
the queue size.
\begin{lstlisting}
  decl conceal[A]{n} : (q : queue[A]{n}) |- (u : uqueue[A])

  proc u <- conceal[A]{n} q =
    case u (                             % receive label along unsized queue u
     ins => x <- recv q ;                % if `ins', receive channel x along u
             q.ins ;                     % send label `ins' along sized queue q
             send q x ;                  % send channel x along q
             u <- conceal[A]{n+1} q      % continue as conceal at index n+1
   | del => q.del ;                      % if `del', send `del' to q
             case q (                    % case on label received from q
               none => u.none ;          % if `none', send `none' on u
                        u <-> q          % identify and terminate
             | some => x <- recv q ;     % receive channel x from q
                        u.some ;         % send `some' on u
                        send u x ;       % send channel x on u
                        u <- conceal[A]{n-1} q
             )
    )
\end{lstlisting}
The \verb'conceal' process case analyzes on the label received on \verb'u'.
If it receives an \verb'ins' request on \verb'u', it forwards the same on
\verb'q' and recurses at index \verb'n+1'
Similarly, it forwards the \verb'del' request to \verb'q' and forwards on
the response from \verb'q' to \verb'u'.
In the \verb'none' branch, it identifies the channels \verb'u' and \verb'q'
and terminates.
In the \verb'some' branch, it forwards the element \verb'x' received from
\verb'q' on to \verb'u' and recurses at index \verb'n+1'.

We can also define a dual \verb'display' process that takes an unsized
queue as input and provides a sized queue.
\begin{lstlisting}
  decl display[A] : (u : uqueue[A]) |- (q : ?n. queue[A]{n})
\end{lstlisting}
In this case, we use existential quantifier to express that there
exists an \verb'n' such that the unsized queue has size \verb'n'.
The \verb'display' process will first send an index natural number
\verb'n' and then behave as a queue of size \verb'n'.
We omit the process definition which also exists in the file
\verb'tests/queues-quant.rast' in the repository.

\section{Basic System of Session Types}\label{sec:basic}

The underlying system of session types is derived from a Curry-Howard
interpretation~\cite{Caires10concur,Caires16mscs} of intuitionistic linear logic~\cite{Girard87tapsoft}. The key idea is that an intuitionistic linear sequent
$A_1 \; A_2 \; \ldots \; A_n \vdash A$
is interpreted as the interface to a process $P$. We label each of the
antecedents with a channel name $x_i$ and the succedent with channel name $z$.
The $x_i$'s are \emph{channels used by} $P$ and $z$ is the \emph{channel provided by} $P$.
\begin{mathpar}
  (x_1 : A_1), (x_2 : A_2), \ldots, (x_n : A_n) \vdash P :: (z : C)
\end{mathpar}
The resulting judgment formally states that process $P$ provides a service of
session type $C$ along channel $z$, while using the services of session types $A_1,
\ldots, A_n$ provided along channels $x_1, \ldots, x_n$ respectively. All these
channels must be distinct. We often abbreviate the linear antecedent of the
sequent by $\Delta$.
Thus, the formal typing judgment is written as
\[
  \D \vdash_\Sg P :: (x : A)
\]
where $\D$ represents the linear antecedents $x_i : A_i$, $P$ is the process
expression and $x : A$ is the linear succedent.
Additionally, $\Sigma$ is a fixed valid signature containing type and process
definitions (explained later).
Because it is fixed, we elide it from the presentation of the typing rules.
We will extend the typing judgment in the subsequent sections as we introduce
polymorphism (Section~\ref{sec:poly}),
refinements (Section~\ref{sec:refine}), ergometric (Section~\ref{sec:ergo}), and
temporal session types (Section~\ref{sec:temporal}).

\begin{table}[t]
  \begin{tabular}{l l l l l}
  \textbf{Type} & \textbf{Cont.} & \textbf{Process Term} & \textbf{Cont.} & \multicolumn{1}{c}{\textbf{Description}} \\
  \toprule
  $c : \ichoice{\ell : A_\ell}_{\ell \in L}$ & $c : A_k$ & $\esendl{c}{k} \semi P$
  & $P$ & send label $k$ along $c$ \\
  & & $\ecase{c}{\ell}{Q_\ell}_{\ell \in L}$ & $Q_k$ & branch on received label along $c$ \\
  \addlinespace
  $c : \echoice{\ell : A_\ell}_{\ell \in L}$ & $c : A_k$ & $\ecase{c}{\ell}{P_\ell}_{\ell \in L}$
  & $P_k$ & branch on received label along $c$ \\
  & & $\esendl{c}{k} \semi Q$ & $Q$ & send label $k$ along $c$ \\
  \addlinespace
  $c : A \tensor B$ & $c : B$ & $\esendch{c}{w} \semi P$
  & $P$ & send channel $w : A$ along $c$ \\
  & & $\erecvch{c}{y} \semi Q$ & $Q[w/y]$ & receive channel $w : A$ along $c$ \\
  \addlinespace
  $c : A \lolli B$ & $c : B$ & $\erecvch{c}{y} \semi P$
  & $P[w/y]$ & receive channel $w : A$ along $c$ \\
  & & $\esendch{c}{w} \semi Q$ & $Q$ & send channel $w : A$ along $c$ \\
  \addlinespace
  $c : \one$ & --- & $\eclose{c}$
  & --- & send $\mi{close}$ along $c$ \\
  & & $\ewait{c} \semi Q$ & $Q$ & receive $\mi{close}$ along $c$ \\
  \bottomrule
  \end{tabular}
  \caption{Basic session types with operational description}%
  \label{tab:language}
\end{table}

At runtime, a program is represented using a multiset of semantic
objects denoting processes and messages defined as a \emph{configuration}.
\[
  \config \; ::= \; \cdot \mid \config, \config' \mid \proc{c}{P} \mid \msg{c}{M}
\]
We formalize the operational semantics as a system of \emph{multiset rewriting
rules}~\cite{Cervesato09SEM}. We introduce semantic objects $\proc{c}{P}$
and $\msg{c}{M}$ which mean that process $P$ or message $M$ provide
along channel $c$.
A process configuration is a multiset of such objects, where any two
provided channels are distinct.

In this section, we briefly review the structural type formers that constitute
the base fragment of Rast.
The type grammar is defined as
\[
  \begin{array}{lrcl}
    \mbox{Types} & A, B, C & ::= & \ichoice{\ell : A}_{\ell \in L}
    \mid \echoice{\ell : A}_{\ell \in L}
                 \mid A \tensor B \mid A \lolli B \mid \one \mid V
  \end{array}
\]
Table~\ref{tab:language} overviews the types, their associated
process terms, their continuation (both in types and terms) and operational description.
For each type, the first row describes the provider's viewpoint, while
the second row describes the client's matching but dual viewpoint.

\paragraph{\textbf{\textit{Choice Operators}}}

The \emph{internal choice} type constructor
$\ichoice{\ell : A_{\ell}}_{\ell \in L}$ is an $n$-ary labeled
generalization of the additive disjunction $A \ichoiceop B$.
Operationally, it requires the provider of
$x : \ichoice{\ell : A_{\ell}}_{\ell \in L}$ to send a label
label $k \in L$ on channel $x$ and continue to provide
type $A_{k}$. The corresponding process term is written as $(\esendl{x}{k} \semi P)$
where the continuation $P$ provides type $x : A_k$.
Dually, the client must branch based
on the label received on $x$ using the process term
$\ecase{x}{\ell}{Q_\ell}_{\ell \in L}$ where $Q_\ell$ is the continuation
in the $\ell$-th branch.

\begin{mathpar}
  \infer[{\oplus}R]
    {\D \vdash (\esendl{x}{k} \semi P) :: (x : \ichoice{\ell : A_\ell}_{\ell \in L})}
    {(k \in L) & \D \vdash P :: (x : A_k)}
  \and
  \infer[{\oplus}L]
    {\D, (x : \ichoice{\ell : A_\ell}_{\ell \in L}) \vdash
    \ecase{x}{\ell}{Q_\ell}_{\ell \in L} :: (z : C)}
    {(\forall \ell \in L) &
      \D, (x : A_\ell) \vdash Q_\ell :: (z : C)}
\end{mathpar}

Communication is asynchronous, so that the provider
($\esendl{c}{k} \semi P$) sends a message $k$ along $c$ and continues as $P$
without waiting for it to be received. As a technical device to ensure that
consecutive messages on a channel arrive in order, the sender also creates a
fresh continuation channel $c'$ so that the message $k$ is actually represented
as $(\esendl{c}{k} \semi \fwd{c}{c'})$ (read: send $k$ along $c$ and continue along
$c'$).  This formulation has the advantage that a message is just a special
form of process only with a different semantic symbol.
The reason we distinguish processes and messages in the semantics is that
messages, unlike processes, are only allowed to interact with other processes,
not spontaneously create messages.
For instance, in the ${\oplus}S$ rule below, if we used $\m{proc}$ to
represent the message, we could apply the same rule recursively to create new
messages.
When the message $k$ is received along $c$, we select branch $k$ and
also substitute the continuation channel $c'$ for $c$.
Rules ${\oplus}S$ and ${\oplus}C$ below describe the operational behavior of the
provider and client respectively.

\begin{tabbing}
$({\oplus}S) :$ \= $\m{proc}(c, c.k \semi P) \;\mapsto\;
\m{proc}(c', P[c'/c]),\; \m{msg}(c, c.k \semi \fwd{c}{c'})$ \qquad $\fresh{c'}$ \\[0.2em]
$({\oplus}C):$ \> $\m{msg}(c, c.k \semi \fwd{c}{c'}),\;
\m{proc}(d, \m{case}\;c\;(\ell \Rightarrow Q_\ell)_{\ell \in L})
\;\mapsto\; \m{proc}(d, Q_k[c'/c])$
\end{tabbing}

The \emph{external choice} constructor $\echoice{\ell : A_{\ell}}_{\ell \in L}$
generalizes additive conjunction and is the \emph{dual} of internal
choice reversing the role of the provider and client. Thus, the provider
branches on the label $k \in L$ sent by the client. The typing rules are as follows
\begin{mathpar}
\infer[{\with}R]
{\D \vdash \ecase{x}{\ell}{P_\ell}_{\ell \in L} ::
(x : \echoice{\ell : A_\ell}_{\ell \in L})}
{(\forall \ell \in L)
 & \D \vdash P_\ell :: (x : A_\ell)}
\and
\infer[{\with}L]
{\D, (x : \echoice{\ell : A_\ell}_{\ell \in L}) \vdash
(\esendl{x}{k} \semi Q) :: (z : C)}
{(k \in L) & \D, (x : A_k) \vdash Q :: (z : C)}
\end{mathpar}
Semantics rules ${\with}S$ and ${\with}C$ express the operational behavior at runtime.
\begin{tabbing}
$({\with}S):$ \= $\m{proc}(d, c.k \semi Q) \;\mapsto\; \m{msg}(c', c.k \semi \fwd{c'}{c}),\;
\m{proc}(d, Q[c'/c]) \qquad \fresh{c'}$ \\[0.2em]
$({\with}C):$ \> $\m{proc}(c, \m{case}\;c\;(\ell \Rightarrow Q_\ell)_{\ell \in L}),\;
\m{msg}(c', c.k \semi \fwd{c'}{c})
\;\mapsto\; \m{proc}(c', Q_k[c'/c])$
\end{tabbing}

\paragraph{\textbf{\textit{Channel Passing}}}

The \emph{tensor} operator $A \tensor B$ prescribes that the provider of
$x : A \tensor B$
sends a channel, say $w$ of type $A$, and continues to provide type $B$. The
corresponding process term is $(\esendch{x}{w} \semi P)$ where $P$ is
the continuation.  Correspondingly, its client must receive a channel on $x$
using the term $(\erecvch{x}{y} \semi Q)$, binding it to variable $y$
and continuing to execute $Q$.
\begin{mathpar}
  \infer[{\tensor}R]
    {\D, (y : A) \vdash (\esendch{x}{y} \semi P) :: (x : A \tensor B)}
    {\D \vdash P :: (x : B)}
  \and
  \infer[{\tensor}L]
    {\D, (x : A \tensor B) \vdash (\erecvch{x}{y} \semi Q) :: (z : C)}
    {\D, (y : A), (x : B) \vdash Q :: (z : C)}
\end{mathpar}
Operationally, the provider $(\esendch{c}{d} \semi P)$ sends the
channel $d$ and the continuation channel $c'$ along $c$ as a message and
continues with executing $P$. The client receives the channel $d$ and continuation
channel $c'$ and substitutes $d$ for $x$ and $c'$ for $c$.
\begin{tabbing}
$({\tensor}S):$ \= $\m{proc}(c, \m{send}\; c\; d \semi P) \;\mapsto\; $\=
$\m{proc}(c', P[c'/c]),\;
\m{msg}(c, \m{send}\; c\; d \semi \fwd{c}{c'}) \qquad \fresh{c'}$ \\[0.2em]
$({\tensor}C):$ \> $\m{msg}(c, \m{send}\; c\; d \semi \fwd{c}{c'}),\;
\m{proc}(e, x \leftarrow \m{recv}\; c \semi Q)
\;\mapsto\; \m{proc}(e, Q[c', d/c, x])$
\end{tabbing}

The dual operator $A \lolli B$ allows the provider $(\erecvch{x}{y} \semi P)$
to receive a channel of type $A$ and continue to provide type $B$ with
process term $P$. The client
of $A \lolli B$, on the other hand, sends channel $w$ of type $A$
and continues to use $B$ using term $(\esendch{x}{w} \semi Q)$.
\begin{mathpar}
  \infer[{\lolli}R]
    {\D \vdash (\erecvch{x}{y} \semi P) :: (x : A \lolli B)}
    {\D, (y : A) \vdash P :: (x : B)}
  \and
  \infer[{\lolli}L]
    {\D, (x : A \lolli B), (y : A) \vdash (\esendch{x}{y} \semi Q) :: (z : C)}
    {\D, (x : B) \vdash Q :: (z : C)}
\end{mathpar}
The semantics rules are the exact dual to $\tensor$.
\begin{tabbing}
  $({\lolli}S) :$ \= $\m{proc}(e, \m{send}\; c\; d \semi Q) \;\mapsto\;
  \m{msg}(c', \m{send}\; c\; d \semi \fwd{c'}{c}),\;
  \m{proc}(e, Q[c'/c]) \qquad \fresh{c'}$ \\[0.2em]
  $({\lolli}C) :$ \> $\m{proc}(c, x \leftarrow \m{recv}\; c \semi P),\;
  \m{msg}(c', \m{send}\; c\; d \semi \fwd{c'}{c})
  \;\mapsto\; \m{proc}(c', P[c', d/c, x])$
\end{tabbing}

\paragraph{\textbf{\textit{Termination}}}

The type $\one$
indicates \emph{termination} requiring that the provider of $x : \one$
send a \emph{close} message, formally written as $(\eclose{x})$
followed by terminating the communication. Correspondingly,
the client of $x : \one$ uses the term $(\ewait{x} \semi Q)$
to wait for the close message before continuing with executing $Q$.
Linearity enforces that the provider does not use any channels,
as indicated by the empty context in ${\one}R$.
\begin{mathpar}
  \infer[{\one}R]
    {\cdot \vdash (\eclose{x}) :: (x : \one)}
    { }
  \and
  \infer[{\one}L]
    {\D, (x : \one) \vdash (\ewait{x} \semi Q) :: (z : C)}
    {\D \vdash Q :: (z : C)}
\end{mathpar}
Operationally, the provider waits for the closing message, which
has no continuation channel since the provider terminates.
\begin{tabbing}
$({\one}S) :$ \= $\m{proc}(c, \m{close}\; c) \;\mapsto\; \m{msg}(c, \m{close}\; c)$ \\[0.2em]
$({\one}C) :$ \> $\m{msg}(c, \m{close}\; c),\;
\m{proc}(d, \m{wait}\; c \semi Q) \;\mapsto\; \m{proc}(d, Q)$
\end{tabbing}

\paragraph{\textbf{\textit{Forwarding}}}
A forwarding process $(\fwd{x}{y})$ identifies the channels $x$ and $y$ so that any
further communication along either $x$ or $y$ will be along the unified channel.
Its typing rule corresponds to the logical rule of identity.
\begin{mathpar}
  \infer[\m{id}]
    {y : A \vdash (\fwd{x}{y}) :: (x : A)}
    {\mathstrut}
\end{mathpar}
Operationally, a process $\fwd{c}{d}$ \emph{forwards} any message M
that arrives on $d$ to $c$ and vice-versa. Since channels are used
linearly, the forwarding process can then terminate, ensuring proper
renaming, as exemplified in the rules below.
\begin{tabbing}
$(\m{id}^+C) : $ \= $\m{msg}(d, M),\;
\m{proc}(c, \fwd{c}{d}) \;\mapsto\; \m{msg}(c, M[c/d])$ \\
$(\m{id}^-C) : $ \> $\m{proc}(c, \fwd{c}{d}),\;
\m{msg}(e, M(c)) \;\mapsto\; \m{msg}(e, M(c)[d/c])$
\end{tabbing}
$M(c)$ indicates that $c$ occurs in message $M$ ensuring that $M$ is the
sole client of $c$.

\paragraph{\textbf{\textit{Process Definitions}}}

Process definitions have the form
$\D \vdash f = P :: (x : A)$ where $f$ is the name of the
process and $P$ its definition, with $\D$ being the channels used
by $f$ and $x : A$ being the provided channel.
All definitions are collected in a fixed global signature $\Sg$.
For a \emph{valid signature}, we
require that $\D \vdash P :: (x : A)$
for every definition, thereby allowing
definitions to be mutually recursive. A new instance of a defined
process $f$ can be spawned with the expression
$\ecut{x}{f}{\overline{y}}{Q}$ where $\overline{y}$ is a
sequence of channels matching the antecedents $\D$.
The newly spawned process will use all variables in
$\overline{y}$ and provide $x$ to the continuation $Q$.
The following $\m{def}$ rule describes the typing of a spawn.

\begin{mathpar}
  \inferrule*[right=$\m{def}$]
  {\overline{y':B'} \vdash f = P_f :: (x' : B) \in \Sg \and
  \D' = \overline{(y:B')} \and
  \D, (x : B) \vdash Q :: (z : C)}
  {\D, \D' \vdash_\Sg (\ecut{x}{f}{\overline{y}}{Q}) :: (z : C)}
\end{mathpar}
The declaration of $f$ is looked up in the signature $\Sg$ (first premise),
and the channel types in $\D$ are matched to the signature $\overline{B'}$
(second premise). Similarly, the freshly created channel $x$ has
type $B$ from the signature.
The corresponding semantics rule $\m{def}C$ also performs a similar substitution.
\begin{tabbing}
$(\m{def}C) : $ \= $\m{proc}(c, \ecut{x}{f}{\overline{d}}{Q}) \; \mapsto \;
\m{proc}(a, P_f[a,\overline{d}/x',\overline{y'}]), \;
   \m{proc}(c, Q[a/x]) \qquad \fresh{a}$
\end{tabbing}
where $\overline{y' : B'} \vdash f = P_f :: (x' : B) \in \Sg$.

Sometimes a process invocation is a tail call,
written without a continuation as $\procdef{f}{\overline{y}}{x}$. This is a
short-hand for $\procdef{f}{\overline{y}}{x'} \semi \fwd{x}{x'}$ for a fresh
variable $x'$, that is, we create a fresh channel
and immediately identify it with x.

\paragraph{\textbf{\textit{Type Definitions}}}
Session types can be defined
recursively, departing from a strict Curry-Howard interpretation of
linear logic, analogous to the way pure ML or Haskell depart from
a pure interpretation of intuitionistic logic.
A type definition for a type name $V$ is of the form
$V = A$, where $A$ is a type expression.
The signature $\Sg$ contains all type definitions,
that are possibly mutually recursive.
For a well-formed signature, we require $A$ to be
\emph{contractive}, i.e., $A$ itself must not be a type name.
Our type definitions are \emph{equirecursive} so we can
silently replace type names $V$ by $A$ during type checking,
and no explicit rules for recursive types are needed.
Because both process definitions and type definitions may be
recursive, processes in our language may not be terminating.

\section{Polymorphic Session Types}\label{sec:poly}

In this section, we describe the modifications to the Rast language
required to realize nested polymorphism~\cite{Das21ESOP} to support general-purpose programming.
First, due to the presence of type variables, the formal typing judgment
is extended with $\vars$ and written as
\[
  \vars \semi \D \vdash_\Sg P :: (x : A)
\]
where $\vars$ stores the type variables (which we denote by $\alpha$).
We presuppose and maintain that all free type variables in $\D, P$, and $A$
are contained in $\vars$. The signature $\Sigma$ is still fixed but the
process and type definitions are now (possibly) parameterized by type variables.

To support polymorphism, we need two primary additions. First, we need to
update the form of process
and type definitions. Secondly, we add explicit quantifiers to allow exchange
of types at runtime. Thus, the extended type grammar is
\[
  \begin{array}{lrcl}
    \mbox{Types} & A & ::= & \ldots \mid \alpha \mid V \indv{A}
    \mid \texists{\alpha} A \mid \tforall{\alpha} A
  \end{array}
\]

\paragraph{\textbf{\textit{Type Definitions}}}
A type name $V$ is now defined according to a definition
$V\indv{\alpha} = A$ that is parameterized by a sequence of
\emph{distinct type variables} $\overline\alpha$ that its definition $A$ can refer to.
We can use type names in an expression using $V \indv{B}$.
Type expressions can also refer to parameter $\alpha$
available in scope.
The \emph{free variables} in type $A$ refer to the set of type variables
that occur freely in $A$.
Since types are \emph{equirecursive}, the type $V \indv{B}$ is considered
equivalent to its unfolding $A[\overline{B}/\overline{\alpha}]$.
All type names $V$ occurring in a valid signature
must be defined, and all type variables defined in a valid
definition must be distinct.
Furthermore, for a valid definition $V \indv{\alpha} = A$, the free variables occurring
in $A$ must be contained in $\overline{\alpha}$.

\paragraph{\textbf{\textit{Process Definitions}}}
Process definitions now have the form
$\D \vdash f\indv{\alpha} = P :: (x : A)$ where $f$ is the name of the
process and $P$ its definition, with $\D$ being the channels used
by $f$ and $x : A$ being the offered channel.
In addition, $\overline{\alpha}$
is a sequence of type variables that $\D$, $P$ and $A$ can refer to.
These type variables are implicitly universally quantified at the
outermost level.
The spawn expression $(\ecut{x}{f \indv{A}}{\overline{y}}{Q})$
now takes a sequence of types $\overline{A}$ matching the type
variables $\overline{\alpha}$, making the polymorphism explicit.
The $\m{def}$ rule is updated to reflect this modification.
\begin{mathpar}
  \inferrule*[right=$\m{def}$]
  {\overline{y':B'} \vdash f \indv{\alpha} = P_f :: (x' : B) \in \Sg \\\\
  \D' = \overline{(y:B')}[\overline{A}/\overline{\alpha}] \and
  \vars \semi \D, (x : B[\overline{A}/\overline{\alpha}]) \vdash Q :: (z : C)}
  {\vars \semi \D, \D' \vdash (\ecut{x}{f \indv{A}}{\overline{y}}{Q}) :: (z : C)}
\end{mathpar}
Note that $\overline{A}$ is substituted for $\overline{\alpha}$ while
matching the types in $\D'$ and $\overline{y}$ (second premise). Similarly,
the provided channel $x$ has type $B$ from the signature
with $\overline{A}$ substituted for $\overline{\alpha}$.

\paragraph{\textbf{\textit{Explicit Quantifiers}}}

To support full parametric polymorphism, Rast also provides
two dual type constructors, $\texists{\alpha} A$
and $\tforall{\alpha} A$ to exchange types between processes.
Table~\ref{tab:quant} describes these types along with
their operational behavior.
The existential type $\texists{\alpha}A$ requires the provider
to send a valid type $B$ and  continue to provide $A[B/\alpha]$. The
term used to send a type is $(\esendch{x}{[B]} \semi P)$ where $P$
is the continuation. Dually, the client receives a type using the
expression $(\erecvch{x}{[\alpha]} \semi Q)$, binds it to variable
$\alpha$, and then continues to execute $Q$ using the session type $A$
which possibly contains the free variable $\alpha$.
We introduce an auxiliary judgment $\vars \vdash \valid{B}$ to
define that all free variables of $B$ are contained in $\vars$.
The corresponding typing rules are

\begin{mathpar}
  \infer[{\exists}R]
    {\vars \semi \D \vdash (\esendch{x}{[B]} \semi P) :: (x : \texists{\alpha}A)}
    { \vars \vdash \valid{B} \and
    \vars \semi \D \vdash P :: (x : A[B/\alpha])}
  \and
  \infer[{\exists}L]
  	{\vars \semi \D, (x : \texists{\alpha}A ) \vdash (\erecvch{x}{[\alpha]} \semi Q) :: (z : C)}
    { \vars, \alpha \semi \D, (x : A) \vdash Q :: (z : C)}
\end{mathpar}
The provider checks whether type $B$ is valid and continues to provide
type $A[B/\alpha]$. The client receives a type, binding it to $\alpha$
which is ensured by adding $\alpha$ to $\vars$.

Operationally, the provider $(\esendch{x}{[B]} \semi P)$ sends the
type $B$ and the continuation channel $c'$ along $c$ and
continues executing $P$. The client receives the type $B$ and continuation
channel $c'$ and substitutes $B$ for $\alpha$ and $c'$ for $c$ in Q.
\begin{tabbing}
$({\exists}S) :$ \= $\m{proc}(c, \m{send}\; c\; [B] \semi P) \;\mapsto\; $\=
$\m{proc}(c', P[c'/c]),\;
\m{msg}(c, \m{send}\; c\; [B] \semi \fwd{c}{c'}) \qquad \fresh{c'}$ \\[0.2em]
$({\exists}C) : $ \> $\m{msg}(c, \m{send}\; c\; [B] \semi \fwd{c}{c'}),\;
\m{proc}(e, [\alpha] \leftarrow \m{recv}\; c \semi Q)
\;\mapsto\; \m{proc}(e, Q[c', B/c, \alpha])$
\end{tabbing}

\noindent
Dually, a provider with a universally typed session $\tforall{\alpha}A$
receives an arbitrary type, binds it to $\alpha$, and proceeds to provide
session prescribed by type $A$, possibly referring $\alpha$.
On the other hand, the client sends a valid type $B$ and
continues with session $A[B/\alpha]$. The formal rules for $\tforall{\alpha}A$ are
\begin{mathpar}
  \infer[{\forall}R]
  	{\vars \semi \D \vdash (\erecvch{x}{[\alpha]} \semi P) :: (x : \tforall{\alpha}A)}
    {\vars, \alpha \semi \D \vdash P :: (x : A)}
  \and
  \infer[{\forall}L]
  	{\vars \semi \D, (x : \tforall{\alpha}A ) \vdash (\esendch{x}{[B]} \semi Q) :: (z : C)}
    { \vars \vdash \valid{B} \and
    \vars \semi \D, (x : A [B/\alpha]) \vdash Q :: (z : C)}
\end{mathpar}

\begin{tabbing}
  $({\forall}S) :$ \= $\m{proc}(d, \m{send}\; c\; [B] \semi P) \;\mapsto\;
  \m{msg}(c', \m{send}\; c\; [B] \semi \fwd{c'}{c}),\;
  \m{proc}(d, P[c'/c]) \qquad \fresh{c'}$ \\[0.2em]
  $({\forall}C) : $ \> $\m{proc}(c, [\alpha] \leftarrow \m{recv}\; c \semi Q),\;
  \m{msg}(c', \m{send}\; c\; [B] \semi \fwd{c'}{c})
  \;\mapsto\; \m{proc}(c', Q[c', B/c, \alpha])$
\end{tabbing}

\noindent
Since polymorphism is parametric, it is possible to avoid explicitly
sending types at runtime if this optimization is desired and does not
interfere with other lower-level aspects of an implementation such as
dynamic monitoring.

\begin{table}[t]
  \small
  \begin{tabular}{l l l l l}
  \textbf{Type} & \textbf{Cont.} & \textbf{Process Term} & \textbf{Cont.} & \multicolumn{1}{c}{\textbf{Description}} \\
  \toprule
  $c : \texists{\alpha}A $ & $c : A[B/\alpha]$ & $\esendch{c}{[B]} \semi P$
  & $P$ & provider sends type $B$ along $c$ \\
  & & $\erecvch{c}{[\alpha]} \semi Q_{\alpha}$ & $Q_{\alpha}[B/{\alpha}]$ & client receives type $B$ along $c$ \\
  \addlinespace
  $c : \tforall{\alpha}A $ & $c : A[B/\alpha]$ & $\erecvch{c}{[\alpha]} \semi P_{\alpha}$
  & $P_{\alpha}[B/\alpha]$ & provider receives type $B$ along $c$ \\
  & & $\esendch{c}{[B]} \semi Q$ & $Q$ & client sends type $B$ along $c$ \\
  \bottomrule
  \end{tabular}
  \caption{Explicitly quantified session types with operational description}%
  \label{tab:quant}
\end{table}

\paragraph{\textbf{\textit{Example: Context-Free Languages}}}
Recursive session types
capture the class of regular languages~\cite{Thiemann16icfp}.
However, in practice, many useful languages are beyond regular.
As an illustration, suppose we would like to express a balanced parentheses
language, also known as the Dyck language~\cite{Dyck1882} with the
end-marker $\$$.
We use $\mb{L}$ to denote an opening
symbol, and $\mb{R}$ to denote a closing symbol
(in a session-typed mindset, $\mb{L}$ can represent client request
and $\mb{R}$ is server response). We need to enforce
that each $\mb{L}$ has a corresponding closing $\mb{R}$ and they are
properly nested.
To express this, we need to track the
number of $\mb{L}$'s in the output with the session type. However,
this notion of \emph{memory} is beyond the expressive power of regular languages,
so mere recursive session types will not suffice.

We utilize the expressive power of nested types to express this behavior.
\begin{lstlisting}
  type T[x] = +{L : T[T[x]], R : x}
  type D = +{L : T[D], $ : 1}
\end{lstlisting}
The nested type \verb'T[T[x]]' takes \verb'x' as a type parameter and either outputs
$\mb{L}$ and continues with \verb'T[T[x]]', or outputs $\mb{R}$ and continues
with \verb'x'. The type \verb'D' either outputs $\mb{L}$ and continues
with \verb'T[D]', or outputs $\$$ and terminates. The type \verb'D' expresses
a Dyck word~\cite{korenjak1966simple}.

The key idea here is that the number of \verb'T''s in the type of a word
tracks the number of unmatched $\mb{L}$'s in it. Whenever the type \verb'T[x]'
outputs $\mb{L}$, it recurses with \verb'T[T[x]]' incrementing the number of
\verb'T''s in the type by $1$. Dually, whenever the type outputs $\mb{R}$, it
recurses with $x$ decrementing the number of \verb'T''s in the type by \verb'1'.
The type \verb'D' denotes a balanced word with no unmatched $\mb{L}$'s.
Moreover, since we can only output $\$$
(or $\mb{L}$) at the type \verb'D' and \emph{not} $\mb{R}$, we obtain the invariant that
any word of type \verb'D' must be balanced.
If we imagine the parameter $x$ as the symbol stack, outputting an $\mb{L}$
pushes \verb'T' on the stack, while outputting $\mb{R}$ pops \verb'T' from the stack.
The definition of \verb'D' ensures that once an $\mb{L}$ is outputted, the symbol
stack is initialized with \verb'T[D]' indicating one unmatched $\mb{L}$.
The file \verb'polytests/dyck.rast' in the repository contains the complete code.

\section{Refinement Session Types}\label{sec:refine}

In this section, we index types with arithmetic refinements that describe
intrinsic attributes of corresponding channels.
In addition to the type constructors arising
from the connectives of intuitionistic linear logic ($\oplus$,
$\with$, $\tensor$, $\one$, $\lolli$) and type variables arising
from polymorphism, we index type names by a
sequence of arithmetic expressions $V \indv{A} \indn{e}$, existential and
universal quantification over natural numbers ($\texists{n} A$,
$\tforall{n} A$) and existential and universal constraints
($\tassert{\phi} A$, $\tassume{\phi} A$).
The type grammar is extended as follows (we write $i$ for constant
and $n$ for variable natural numbers).
\[
  \begin{array}{lrcl}
    \mbox{Types} & A, B, C & ::= & \ldots \mid V \indv{A} \indn{e}
                    \mid \tassert{\phi} A \mid \tassume{\phi} A
                            \mid \texists{n} A \mid \tforall{n} A \\[0.8em]
    \mbox{Arith. Exps.} & e & ::= & i \mid e + e \mid e - e \mid i \times e \mid n \\[0.8em]
    \mbox{Arith. Props.} &
    \phi & ::= & e = e \mid e > e \mid \top \mid \bot
                 \mid \phi \land \phi \\
    & & \mid &  \phi \lor \phi \mid \lnot \phi \mid \texists{n}\phi \mid \tforall{n} \phi
  \end{array}
\]

To account for refinements, the typing judgment is further extended and has the form
\[
  \Vars \semi \cons \semi \vars \semi \D \vdash_\Sg P :: (x : A)
\]
where $\Vars$ are arithmetic variables $n$, $\cons$ are constraints over
these variables expressed as a single proposition,
$\D$ are the linear antecedents $x_i : A_i$, $P$ is a process
expression, and $x : A$ is the linear succedent. We presuppose and maintain
that all free index variables in $\cons$, $\D$, $P$, and $A$ are contained among $\Vars$.
As described in Section~\ref{sec:poly}, the process and type definitions in signature $\Sg$
are now also parameterized by arithmetic variables.
In addition we
write $\Vars \semi \cons \proves \phi$ for semantic entailment
(proving $\phi$ assuming $\cons$) in the
constraint domain where $\Vars$ contains all arithmetic variables in
$\cons$ and $\phi$.

We now describe quantifiers ($\texists{n} A$, $\tforall{n} A$) and
constraints ($\tassert{\phi} A$, $\tassume{\phi} A$).  An overview
of the types, process expressions, their continuation, and operational meaning can
be found in Table~\ref{tab:refine}.

\begin{table}[t]
  \begin{tabular}{l l l l l}
  \textbf{Type} & \textbf{Cont.} & \textbf{Process Term} & \textbf{Cont.} & \multicolumn{1}{c}{\textbf{Description}} \\
  \toprule
  $c : \texists{n} A$ & $c : A[i/n]$ & $\esendn{c}{e} \semi P$
  & $P$ & provider sends value $i$ of $e$ along $c$ \\
  & & $\erecvn{c}{n} \semi Q$ & $Q[i/n]$ & client receives number $i$ along $c$ \\
  \addlinespace
  $c : \tforall{n} A$ & $c : A[i/n]$ & $\erecvn{c}{n} \semi P$
  & $P[i/n]$ & provider receives number $i$ along $c$ \\
  & & $\esendn{c}{e} \semi Q$ & $Q$ & client sends value $i$ of $e$ along $c$ \\
  \addlinespace
  $c : \tassert{\phi} A$ & $c : A$ & $\eassert{c}{\phi} \semi P$
  & $P$ & provider asserts $\phi$ on channel $c$ \\
  & & $\eassume{c}{\phi} \semi Q$ & $Q$ & client assumes $\phi$ on $c$ \\
  \addlinespace
  $c : \tassume{\phi} A$ & $c : A$ & $\eassume{c}{\phi} \semi P$
  & $P$ & provider assumes $\phi$ on channel $c$ \\
  & & $\eassert{c}{\phi} \semi Q$ & $Q$ & client asserts $\phi$ on $c$ \\
  \bottomrule
  \end{tabular}
  \caption{Refined session types with operational description}%
  \label{tab:refine}
\end{table}

\paragraph{\textbf{\textit{Quantification}}}
The provider of $(c : \texists{n} A)$ should send a witness $i$ along
channel $c$ and then continue as $A[i/n]$.  The witness is specified
by an arithmetic expression $e$ which, since it must be closed at
runtime, can be evaluated to a number $i$ (following standard evaluation
rules of arithmetic). From the typing perspective,
we just need to check that the expression $e$ denotes a natural
number, using only the permitted variables in $\Vars$.  This is
represented with the auxiliary judgment $\Vars \semi \cons \vdash e : \m{nat}$
(implicitly proving $e \geq 0$ under constraint $\cons$).
\begin{mathpar}
  \infer[\exists R]
  {\Vars \semi \cons \semi \vars \semi \D \vdash \esendn{x}{e} \semi P :: (x : \texists{n} A)}
  {\Vars \semi \cons \vdash e : \m{nat} \and
  \Vars \semi \cons \semi \vars \semi \D \vdash P :: (x : A[e/n])}
  \and
  \infer[\exists L]
  {\Vars \semi \cons \semi \vars \semi \D, (x : \texists{n} A) \vdash \erecvn{x}{n} \semi Q :: (z : C)}
  {\Vars,n \semi \cons \semi \vars \semi \D, (x : A) \vdash Q :: (z : C) 
  }
\end{mathpar}
Statically, the ${\exists}L$ rule adds $n$ to $\Vars$ ensuring that continuations $Q$ and $A$ can
refer to variables in $\Vars, n$.
Operationally, the provider sends the arithmetic expression with the continuation
channel as a message that the client receives and appropriately substitutes.
\begin{tabbing}
  $({\exists}S) : $ \= $\proc{c}{\esendn{c}{e} \semi P} \;\mapsto\;
  \proc{c'}{P[c'/c]}, \; \msg{c}{ \esendn{c}{e} \semi \fwd{c}{c'}} \qquad \fresh{c'}$ \\[0.2em]
  $({\exists}C) : $ \> $\msg{c}{\esendn{c}{e} \semi \fwd{c}{c'}}, \;
  \m{proc}(d, \erecvn{c}{n} \semi Q) \;\mapsto\;$
  $\proc{d}{Q[e, c'/n, c]}$
\end{tabbing}

\noindent
The dual type $\tforall{n} A$ reverses the role of the provider and
client.  The client sends (the value of) an arithmetic expression $e$
which the provider receives and binds to $n$.

\begin{mathpar}
  \infer[\forall R]
  {\Vars \semi \cons \semi \vars \semi \D \vdash \erecvn{x}{n} \semi P_n :: (x : \tforall{n} A)}
  {\Vars,n \semi \cons \semi \vars \semi \D \vdash P_n :: (x : A)}
  \and
  \infer[\forall L]
  {\Vars \semi \cons \semi \vars \semi \D, (x : \tforall{n} A) \vdash \esendn{x}{e} \semi Q :: (z : C)}
  {\Vars \semi \cons \vdash e : \m{nat} \and
  \Vars \semi \cons \semi \vars \semi \D, (x : A[e/n]) \vdash Q :: (z : C)}
\end{mathpar}
\begin{tabbing}
  $({\forall}S) :$ \= $\proc{d}{\esendn{c}{e} \semi P} \;\mapsto\;$
  $\m{msg}(c', \esendn{c}{e} \semi \fwd{c'}{c}), \; \proc{d}{P[c'/c]} \qquad \fresh{c'}$ \\[0.2em]
  $({\forall}C) :$ \> $\proc{d}{\erecvn{c}{n} \semi Q},\;
  \m{msg}(c', \esendn{c}{e} \semi \fwd{c'}{c}) \mapsto$
  $\proc{d}{Q[e, c'/n, c]}$
\end{tabbing}

\paragraph{\textbf{\textit{Constraints}}}
Refined session types also allow constraints over index variables.  As
we have already seen in the examples, these critically govern
permissible messages.
From the message-passing perspective, the provider of
$(c : \tassert{\phi} A)$ should send a proof of $\phi$ along $c$ and
the client should receive such a proof.  However, since the index
domain is decidable and future computation cannot depend on the form
of the proof (what is known in type theory as \emph{proof
  irrelevance}) such messages are not actually exchanged.  Instead, it
is the provider's responsibility to ensure that $\phi$ holds, while the
client is permitted to assume that $\phi$ is true.  Therefore,
and in an analogy with imperative languages, we write $\eassert{c}{\phi} \semi P$
for a process that \emph{asserts} $\phi$ for channel $c$ and continues
with $P$, while $\eassume{c}{\phi} \semi Q$ \emph{assumes} $\phi$ and
continues with $Q$.

Thus, the typing rules for this new type constructor are
\begin{mathpar}
  \infer[{\tassertop}R]
  {\Vars \semi \cons \semi \vars \semi \D \vdash \eassert{x}{\phi} \semi P :: (x : \tassert{\phi} A)}
  {\Vars \semi \cons \proves \phi & \Vars \semi \cons \semi \vars \semi \D \vdash P :: (x : A)}
  \and
  \infer[{\tassertop}L]
  {\Vars \semi \cons \semi \vars \semi \D, (x : \tassert{\phi} A) \vdash \eassume{x}{\phi} \semi Q :: (z : C)}
  {\Vars \semi \cons \land \phi \semi \vars \semi \D, (x : A) \vdash Q :: (z : C)}
\end{mathpar}
Notice how the provider must verify the truth of $\phi$ given the
currently known constraints $\cons$ (the premise $\Vars \semi \cons \proves \phi$),
while the client assumes $\phi$ by adding it to $\cons$.

Operationally, the provider creates a message containing the constraint (which simply
evaluates to $\top$) that is received by the client. Since the constraints exchanged
at runtime are always trivial, we can skip the communication entirely for constraints.
However, in the formal semantics we still require communication for uniformity with other type constructors.
\begin{tabbing}
$({\tassertop}S):$ \= $\proc{c}{\eassert{c}{\phi} \semi P} \;\mapsto\;$
$\m{proc}(c', P[c'/c]), \; \msg{c}{\eassert{c}{\phi} \semi \fwd{c}{c'}} \quad \fresh{c'}$ \\[0.2em]
$({\tassertop}C):$ \> $\msg{c}{\eassert{c}{\phi} \semi \fwd{c}{c'}}, \;$
$\m{proc}(d, \eassume{c}{\phi'} \semi Q) \;\mapsto\; \proc{d}{Q[c'/c]}$
\end{tabbing}
In well-typed configurations (which arise from executing well-typed processes)
the constraint $\phi$ in these
rules will always be closed and true so there is no need to check this
explicitly.

The dual operator $\tassume{\phi} A$ reverses the role of provider and
client. The provider of $x : \tassume{\phi} A$ may assume the truth of
$\phi$, while the client must verify it.  The dual rules are
\begin{mathpar}
  \infer[{\tassumeop}R]
  {\Vars \semi \cons \semi \vars \semi \D \vdash \eassume{x}{\phi} \semi P :: (x : \tassume{\phi} A)}
  {\Vars \semi \cons \land \phi \semi \vars \semi \D \vdash P :: (x : A)}
  \and
  \infer[{\tassumeop}L]
  {\Vars \semi \cons \semi \vars \semi \D, (x : \tassume{\phi} A) \vdash \eassert{x}{\phi} \semi Q :: (z : C)}
  {\Vars \semi \cons \proves \phi & \Vars \semi \cons \semi \vars \semi \D, (x : A) \vdash Q :: (z : C)}
\end{mathpar}
\begin{tabbing}
$({\tassumeop}S):$ \= $\proc{d}{\eassert{c}{\phi} \semi P} \;\mapsto\;$
$\m{msg}(c', \eassert{c}{\phi} \semi \fwd{c'}{c}), \; \proc{d}{P[c'/c]} \quad \fresh{c'}$ \\[0.2em]
$({\tassumeop}C)$ \> $\proc{d}{w, \eassume{c}{\phi'} \semi Q},\;
\m{msg}(c', \eassert{c}{\phi} \semi \fwd{c'}{c}) \;\mapsto\; \proc{d}{Q[c'/c]}$
\end{tabbing}

\noindent
The remaining issue is how to type-check a branch that is impossible
due to unsatisfiable constraints.  For example, if a client sends
a $\mb{del}$ request to a provider along $c : \m{queue}[A]\{0\}$, the
type then becomes
\[
  c : \ichoice{\mb{none} : \tassert{0 = 0} \one,
  \mb{some} : \tassert{0 = 0} A \tensor \m{queue}[A]\{0 - 1\}}
\]
The client would have to branch on the label received
and then assume the constraint asserted by the provider
\begin{sill}
  $\m{case}\; c\;$ \= $(\, \mb{none} \Rightarrow \eassume{c}{0 = 0} \semi P_1$ \\
  \> $\mid \mb{some} \Rightarrow \eassume{c}{0 > 0} \semi P_2)$
\end{sill}
but what could we write for $P_2$ in the $\mb{some}$ branch?
Intuitively, computation should never get there because the provider
can not assert $0 > 0$.  Formally, we use the process expression
`$\eimposs$' to indicate that computation can never reach this spot:
\begin{sill}
  $\m{case}\; c\;$ \= $(\, \mb{none} \Rightarrow \eassume{c}{0 = 0} \semi P_1$ \\
  \> $\mid \mb{some} \Rightarrow \eassume{c}{0 > 0} \semi \eimposs)$
\end{sill}
In implicit syntax, we can omit
the $\mb{some}$ branch altogether and it would be reconstructed
in the form shown above.
Abstracting away from this example, the typing rule for impossibility
simply checks that the constraints are indeed unsatisfiable
\begin{mathpar}
  \infer[\m{unsat}]
  {\Vars \semi \cons \semi \D \vdash \eimposs :: (x : A)}
  {\Vars \semi \cons \proves \bot}
\end{mathpar}
There is no operational rule for this scenario since in well-typed configurations
the process expression `$\eimposs$' is dead code and can never be reached.

\paragraph{\textbf{\textit{Example: Binary Numbers}}}

As another example,
consider natural numbers in binary representation.  The idea is that,
for example, the number 13 in binary $(1101)_2$ form is represented as
a sequence of messages
$(\mb{b1}, \mb{b0}, \mb{b1}, \mb{b1}, \mb{e}, \mi{close})$ sent or
received on a given channel with the least significant bit first.
Here $\mb{e}$ represents 0 (the empty sequence of bits), while
$\mb{b0}$ and $\mb{b1}$ represent bits 0 and 1, respectively.
\begin{lstlisting}
  type bin = +{ b0 : bin, b1 : bin, e : 1 }
\end{lstlisting}
We can then index binary numbers with their value.
Because
(linear) arithmetic contains no division operator, we express the type
$\m{bin}\{n\}$ of binary numbers with value $n$ using existential
quantification, with the concrete syntax \verb'?k. A' for
$\exists k.\, A$.
\begin{lstlisting}
  type bin{n} = +{ b0 : ?{n > 0}. ?k. ?{n = 2*k}. bin{k},
                     b1 : ?{n > 0}. ?k. ?{n = 2*k+1}. bin{k},
                      e : ?{n = 0}. 1 }
\end{lstlisting}
The constraint that $n > 0$ in the case of $\mb{b0}$ ensures the
representation is unique and there are no leading zeros; the same
constraint for $\mb{b1}$ is in fact redundant.  The
\verb'examples/arith.rast' contains several examples of processes over binary
numbers like addition, multiplication, predecessor, equality and
conversion to and from numbers in unary form.

\section{Ergometric Session Types}\label{sec:ergo}

An important application of refinement types is complexity
analysis. Prior works on resource-aware session types~\cite{Das18lics,Das18icfp,Das21CSF} crucially rely on arithmetic
refinements to express work and time bounds.  The design principle we followed is that
they should be \emph{conservative} over the basic and indexed session
types, so that previously defined programs and type-checking rules do
not change. In this section, we
review the ergometric type system that computes work intuitively defined as the
total operations executed by the system.

The key idea is that \emph{processes store potential} and
\emph{messages carry potential}. This potential can either be consumed
to perform \emph{work} or exchanged using special messages. The type
system provides the programmer with the flexibility to specify what
constitutes work. Thus, programmers can choose to count the
resource they are interested in, and the type system provides the
corresponding upper bound.  Our current examples assign unit cost to
message sending operations (exempting those for index objects or
potentials themselves) effectively counting the total number of
``real'' messages exchanged during a computation.

\begin{table}[t]
  \begin{tabular}{l l l l l}
  \textbf{Type} & \textbf{Cont.} & \textbf{Process Term} & \textbf{Cont.} & \multicolumn{1}{c}{\textbf{Description}} \\
  \toprule
  $c : \tpaypot{A}{r}$ & $c : A$ & $\epay{c}{r} \semi P$
  & $P$ & provider pays $r$ potential units on channel $c$ \\
  & & $\eget{c}{r} \semi Q$ & $Q$ & client gets $r$ potential units on $c$ \\
  \addlinespace
  $c : \tgetpot{A}{r}$ & $c : A$ & $\eget{c}{r} \semi P$
  & $P$ & provider gets $r$ potential units on channel $c$ \\
  & & $\epay{c}{r} \semi Q$ & $Q$ & client pays $r$ potential units on $c$ \\
  \bottomrule
  \end{tabular}
  \caption{Ergometric session types with operational description}%
  \label{tab:ergo}
\end{table}

Two dual type constructors $\tpaypot{A}{r}$ and $\tgetpot{A}{r}$ are
used to exchange potential.
Table~\ref{tab:ergo} contains a description of these types and
their operational behavior.
The provider of $x : \tpaypot{A}{r}$ must
\emph{pay} $r$ units of potential along $x$ using process term
$(\epay{x}{r} \semi P)$, and continue to provide $A$ by executing
$P$. These $r$ units are deducted from the potential stored inside the
sender. Dually, the client must receive the $r$ units of potential
using the term $(\eget{x}{r} \semi Q)$ and add this to its internal
stored potential. Finally, since processes are allowed to store
potential, the typing judgment records the potential available to a
process above the turnstile
$\Vars \semi \cons \semi \vars \semi \D \entailpot{q}_{\Sg} P :: (x : A)$. We
allow potential $q$ to refer to index variables in $\Vars$ to capture
variable potential.  The typing rules for $\tpaypot{A}{r}$ are
\begin{mathpar}
  \infer[{\paypot}R]
  {\Vars \semi \cons \semi \vars \semi \D \entailpot{q} \epay{x}{r_1} \semi P :: (x : \tpaypot{A}{r_2})}
  {\Vars \semi \cons \proves q \geq r_1 = r_2 \and
  \Vars \semi \cons \semi \vars \semi \D \entailpot{q-r_1} P :: (x : A)}
  \and
  \infer[{\paypot}L]
  {\Vars \semi \cons \semi \vars \semi \D, (x : \tpaypot{A}{r_2}) \entailpot{q} \eget{x}{r_1} \semi Q :: (z : C)}
  {\Vars \semi \cons \proves r_1 = r_2 \and
  \Vars \semi \cons \semi \vars \semi \D, (x : A) \entailpot{q+r_1} Q :: (z : C)}
\end{mathpar}

\noindent
In both cases, we check that the exchanged potential in the expression
and type matches ($r_1 = r_2$), and while paying, we ensure that the
sender has sufficient potential to pay.
We use distinct variables $r_1$ and $r_2$ to illustrate that the process and
type expressions can use syntactically different but semantically
equal annotations (e.g. $r_1 = n+n, r_2 = 2*n$).
On the other hand, the receiver adds the $r_1$ units to its process potential.

We extend the semantic objects with work counters: $\proc{c}{w, P}$ (resp. $\msg{c}{w, M}$)
denotes a process (resp.\ message) providing channel $c$, executing $P$ (resp. $M$) and having
performed work $w$ so far. The semantics rules for $\paypot$ are expressed as follows.
\begin{tabbing}
  $({\paypot}S) :$ \= $\proc{c}{w, \epay{c}{r} \semi P} \;\mapsto\;
  \proc{c'}{w, P[c'/c]}, \; \msg{c}{0, \epay{c}{r} \semi \fwd{c}{c'}} \quad \fresh{c'}$ \\[0.2em]
  $({\paypot}C) :$ \> $\msg{c}{w', \epay{c}{r} \semi \fwd{c}{c'}}, \;$
  $\m{proc}(d, w, \eget{c}{r} \semi Q) \;\mapsto\;\proc{d}{w+w', Q[c'/c]}$
\end{tabbing}
The freshly created message in rule ${\paypot}S$ has not performed any work so far,
therefore has $w = 0$. In the ${\paypot}C$ rule, the work performed by the message
is absorbed by the receiving process.
Thus, the total work done is conserved, and no work is dropped by the system.
Note that even though the message was created with $w = 0$, it can interact with
forwarding processes and absorb the work performed by them.
Thus, when they interact with the receiver process, they may have a different
work annotation $w'$.
We follow the same approach for all the rules of operational semantics so far.

The dual type $\tgetpot{A}{r}$ enables the provider to receive potential
that is sent by its client. Its rules are the exact inverse of $\paypot$.
\begin{mathpar}
  \infer[{\getpot}R]
  {\Vars \semi \cons \semi \vars \semi \D \entailpot{q} \eget{x}{r_1} \semi P :: (x : \tgetpot{A}{r_2})}
  {\vars \semi \cons \proves r_1 = r_2 \and
  \Vars \semi \cons \semi \vars \semi \D \entailpot{q+r_1} P :: (x : A)}
\end{mathpar}
\begin{mathpar}
  \infer[{\getpot}L]
  {\Vars \semi \cons \semi \vars \semi \D, (x : \tgetpot{A}{r_2}) \entailpot{q} \epay{x}{r_1} \semi Q :: (z : C)}
  {\vars {\semi} \cons \proves q \geq r_1 = r_2 \and
  \Vars \semi \cons \semi \vars \semi \D, (x : A) \entailpot{q-r_1} Q :: (z : C)}
\end{mathpar}
\begin{tabbing}
$({\getpot}S) :$ \= $\proc{d}{w, \epay{c}{r} \semi P} \;\mapsto\;
\m{msg}(c', 0, \epay{c}{r} \semi \fwd{c'}{c}), \; \proc{d}{w, P[c'/c]} \quad \fresh{c'}$ \\[0.2em]
$({\getpot}C) :$ \> $\proc{c}{w, \eget{c}{r} \semi Q}, \;
\m{msg}(c', w', \epay{c}{r} \semi \fwd{c'}{c}) \;\mapsto\;$
$\proc{c'}{w+w', Q[c'/c]}$
\end{tabbing}

\noindent
We use a special expression $\ework{r} \semi P$ to \emph{perform
work}.  Usually, work actions are inserted by the Rast compiler
based on a cost model selected by the programmer, such as paying one
erg just before every send operation.  The programmer can also select
a model where all operations are free and manually insert calls to
$\ework{r}$.  An example of this is given in the file
\verb'examples/linlam-reds.rast' that counts the number of reductions necessary
for the evaluation of an expression in the linear $\lambda$-calculus
(described in Section~\ref{sec:examples}).
\begin{mathpar}
  \infer[\m{work}]
  {\Vars \semi \cons \semi \vars \semi \D \entailpot{q} \ework{r} \semi P :: (x : A)}
  {\Vars \semi \cons \proves q \geq r &
  \Vars \semi \cons \semi \vars \semi \D \entailpot{q-r} P :: (x : A)}
\end{mathpar}
At runtime, executing the $\ework{r}$ construct increments the
work counter by $r$.
\begin{tabbing}
  $(\m{w}C) :$ \= $\proc{c}{w, \ework{r} \semi P} \;\mapsto\;
  \proc{c}{w+r, P}$
\end{tabbing}

\noindent
Work is \emph{precise}, that is, before terminating a process must
have 0 potential, which can be achieved by explicitly consuming any
remaining potential.

\paragraph{\textbf{\textit{Example: Ergometric Queue}}}

We have already seen
the ergometric types of queues as a bucket brigade in the
introduction.  We show it now in concrete syntax, where \verb'<{p}|'
receives potential \verb'p'.
\begin{lstlisting}
  type queue[A]{n} = &{ins : <{2*n}| A -o queue[A]{n+1},
                         del : <{2}| +{none : ?{n = 0}. 1,
                                         some : ?{n > 0}. A * queue[A]{n-1}}}

  decl empty[A] : . |- (q : queue[A]{0})
  decl elem[A]{n} : (x : A) (r : queue[A]{n}) |- (q : queue[A]{n+1})
\end{lstlisting}
Interestingly, the exact code of Listing~\ref{queue_impl} will check
against this more informative type (see file
\verb'examples/list-work.rast').  The cost model will insert the
appropriate $\ework{r}$ action and reconstruction will insert the
actions to pay and get potential.

For a queue implemented internally as two stacks we can perform an
amortized analysis.
Briefly, the queue process maintains two lists: one ($\mi{in}$) to
store messages when they are enqueued, and a reversed list
($\mi{out}$) from which they are dequeued.  When the client wishes to
dequeue an element and the $\mi{out}$ list is empty, the provider
reverses the $\mi{in}$ list to serve as the new $\mi{out}$ list.  A
careful analysis shows that if this data structure is used linearly,
both insert and delete have constant amortized time.  More
specifically we obtain the type
\begin{lstlisting}
type queue[A]{n} = &{enq : <{6}| A -o queue[A]{n+1},
                       deq : <{4}| +{none : ?{n = 0}. 1,
                                       some : ?{n > 0}. A * queue[A]{n-1}}}
\end{lstlisting}
The program can be found in the file \verb'examples/list-work.rast' in the
repository.

\section{Temporal Session Types}\label{sec:temporal}

Rast also supports \emph{temporal modalities} \emph{next} ($\Next A$),
\emph{always} ($\Box A$), and \emph{eventually} ($\Diamond A$),
interpreted over a linear model of time.  To model computation time,
we use the syntactic form $\m{delay}$ which advances time by one
tick. A particular cost semantics is specified by taking an ordinary,
non-temporal program and adding delays capturing the intended cost.
For example, if only the blocking operations should cost one unit of
time, a delay is added before the continuation of every receiving
construct.  For type checking, the $\m{delay}$ construct subtracts one
$\Next$ operator from every channel it refers to.  We denote consuming
$t$ units on the left of the context using $\lforce{A}{-t}$, and on the
right by $\rforce{A}{-t}$. Briefly,
$\lforce{\next^t A}{-t} = \rforce{\next^t A}{-t} = A$.
As we will explain soon, consuming time units on the left and the right
differ due to the $\Box$ and $\Dia$ modalities.

\begin{mathpar}
  \infer[{\Next}LR]
  {\Vars \semi \cons \semi \vars \semi \D \entailpot{q} \edelay{t} \semi P :: (x : A)}
  {\Vars \semi \cons \proves t \geq 0 \\
  \Vars \semi \cons \semi \vars \semi \lforce{\D}{-t} \entailpot{q} Q :: (x : \rforce{A}{-t})}
\end{mathpar}

To express the semantics, we now use $\proc{c}{w, t, P}$ to denote
a process $P$ at local clock $t$. This local clock advances by $r$
as the process executes a $\edelay{r}$.
\begin{tabbing}
  $({\Next}C) :$ \= $\proc{c}{w, t, \edelay{r} \semi P} \;\mapsto\;
  \proc{c}{w, t+r, P}$
\end{tabbing}

\paragraph{\textbf{\textit{Always $A$}}}
A process providing $x : \Box A$ promises to be available at any
time in the future, including now.  When the client would like
to use this provider it (conceptually) sends a message $\noww$
along $x$ and then continues to interact according to type $A$.

A process $P$ providing $x : \Box A$ must be able to wait
indefinitely.  But this is only possible if all the channels that $P$
uses can also wait indefinitely.  This is enforced in the rule by the
condition $\delayedbox{\D}$ which requires each antecedent to have the
form $y_i : \Next^{n_i}\, \Box B_i$.
\begin{mathpar}
  \infer[{\Box}R]
    {\D \vdash (\ewhen{x} \semi P) :: (x : \Box A)}
    {\delayedbox{\D}
      & \D \vdash P :: (x : A)}
  \and
  \infer[{\Box}L]
    {\D, x : \Box A \vdash (\enow{x} \semi Q) :: (z : C)}
    {\D, x : A \vdash Q :: (z : C)}
\end{mathpar}
The corresponding semantics rules are
\begin{tabbing}
$({\Box}S) :$ \= $\m{proc}(d, w, t, \enow{c} \semi P) \; \mapsto \;
\m{msg}(c', 0, t, \enow{c} \semi \fwd{c'}{c}),\; \m{proc}(d, w, t, P[c'/c])$ \\[0.2em]
$({\Box}C) :$ \> $\m{proc}(c, w, s, \ewhen{c} \semi Q), \m{msg}(c', w', t, \enow{c} \semi \fwd{c'}{c})
\;\mapsto\;$\\[0.1em]
\hspace{22em}$\m{proc}(c', w+w', t, Q[c'/c]) \qquad (s \leq t)$
\end{tabbing}

\paragraph{\textbf{\textit{Eventually $A$}}}
The dual of $\tbox{A}$ is $\tdia{A}$.
A process providing $\Diamond A$ promises to provide $A$ eventually.
When a process offering $x : \Dia A$ is ready, it will send a $\noww$
message along $x$ and then continue at type $A$. Conversely, the
client of $x : \Dia A$ will have to be ready and waiting for the
$\noww$ message to arrive along $x$ and then continue at type $A$. We
use $(\ewhen{c} \semi Q)$ for the corresponding client.
The typing rules for $\noww$ and $\whenn$ are somewhat subtle.
\begin{mathpar}
  \infer[{\Dia}R]
  {\D \vdash (\enow{x} \semi P) :: (x : \Dia A)}
  {\D \vdash P :: (x : A)}
  \and
\infer[{\Dia}L]
  {\D, x{:}\Dia A \vdash (\ewhen{x} \semi Q) :: (z : C)}
  {\delayedbox{\D}
    & \D, x : A \vdash Q :: (z : C)
    & \delayeddia{A}}
\end{mathpar}
The predicate $\delayeddia{C}$ means that $C$ must have the form
$\Next^n \Dia C'$ (for some $n$) requiring that $C$ may be delayed a fixed
finite number of time steps and then must be allowed to communicate at
an arbitrary time in the future.
The semantics rules for $\Dia$ are exact inverse of $\Box$.
\begin{tabbing}
$({\Dia}S) :$ \= $\m{proc}(c, w, t, \enow{c} \semi P) \; \mapsto \;
\m{proc}(c', w, t, P[c'/c]),\; \m{msg}(c, 0, t, \enow{c} \semi \fwd{c}{c'})$ \\[0.2em]
$({\Dia}C) :$ \> $\m{msg}(c, 0, t, \enow{c} \semi \fwd{c}{c'}),\; \m{proc}(d, w, s, \ewhen{c} \semi P)
\;\mapsto\;$\\[0.1em]
\hspace{22em}$\m{proc}(d, w+w', t, P[c'/c]) \qquad (s \leq t)$
\end{tabbing}

Table~\ref{tab:temporal} provides a formal description of the temporal types.
Since all temporal operators ultimately model time, they interact with
each other and the temporal displacement operator used in $\Next LR$
rule needs to be generalized as described below.  We define $[A]^0 = A$ and
$[A]^{-(t+1)} = [[A]^{-1}]^{-t}$.  Below $S$ denotes a non-temporal
type.  When the displacement is undefined, the $\Next LR$ rule cannot
be applied.
More details can be found in prior work~\cite{Das18icfp}.

\[
\begin{array}{lcl@{\hspace{3em}}lcl@{\hspace{3em}}lcl}
[\Next A]_L^{-1} & = & A & [\Next A]_R^{-1} & = & A
  & [x:A]_L^{-1} & = & x:[A]_L^{-1} \\\relax
[\Box A]_L^{-1} & = & \Box A & [\Box A]_R^{-1} & = & \mbox{\it undefined}
  & [x:A]_R^{-1} & = & x:[A]_R^{-1} \\\relax
[\Dia A]_L^{-1} & = & \mbox{\it undefined} & [\Dia A]_R^{-1} & = & \Dia A
  & [\cdot]_L^{-1} & = & \cdot \\\relax
[S]_L^{-1} & = & \mbox{\it undefined} & [S]_R^{-1} & = & \mbox{\it undefined}
  & [\D, \D']_L^{-1} & = & [\D]_L^{-1}, [\D']_L^{-1}
\end{array}
\]

\begin{table}[t]
  \begin{tabular}{l l l l l}
  \textbf{Type} & \textbf{Cont.} & \textbf{Process Term} & \textbf{Cont.} & \multicolumn{1}{c}{\textbf{Description}} \\
  \toprule
  $c : \tdia{A}$ & $c : A$ & $\enow{c} \semi P$
  & $P$ & provider sends $\m{now!}$ on channel $c$ \\
  & & $\ewhen{c} \semi Q$ & $Q$ & client waits for $\m{now!}$ on $c$ \\
  \addlinespace
  $c : \tbox{A}$ & $c : A$ & $\ewhen{c} \semi P$
  & $P$ & provider waits for $\m{now!}$ on channel $c$ \\
  & & $\enow{c} \semi Q$ & $Q$ & client sends $\m{now!}$ on $c$ \\
  \addlinespace
  $c : \Next^r A$ & $c : A$ & $\edelay{r} \semi P$ & $P$ &
  provider delays for $r$ time units \\
  \bottomrule
  \end{tabular}
  \caption{Temporal session types with operational description}%
  \label{tab:temporal}
\end{table}

\paragraph{\textbf{\textit{Example: Temporal Queue}}}

We have already foreshadowed the temporal type of a queue, implemented
as a bucket brigade.  We show it now in concrete syntax, where
\verb'()' is the $\Next$ modality and \verb'[]' represents $\Box$.
We also show the types of the \verb'empty' and
\verb'elem' processes (see file \verb'examples/time.rast').

\begin{lstlisting}
  type queue[A]{n} = [] {enq : () A -o ()()()queue[A]{n+1},
                            deq : ()+{none: () ?{n = 0}. 1,
                                       some: () ?{n > 0}. A * ()queue[A]{n-1}}}

  decl empty[A] : . |- (q : ()()queue[A]{0})
  decl elem[A]{n} : (x : A) (r : ()()queue[A]{n}) |- (q : queue[A]{n+1})

\end{lstlisting}

Because Rast currently does not have reconstruction for time we have
to update the program with the five temporal actions presented in this
section (two instances of \verb'delay', two of \verb'when', and one of
\verb'now').
A key observation here is that in the case of \verb'elem' the process
$r$ does not need to be ready instantaneously, but can be ready after
a delay of 2 ticks, because that is how long it takes to receive the
$\mb{ins}$ label and the element along $q$.  This slack is also
reflected in the type of \verb'empty' because it becomes then back of
a new element when the end of the queue is reached.

\section{Subtyping}\label{sec:subtyping}

A late addition to Rast was the introduction of subtyping as a
generalization of type equality.  Declaratively, in the system
of basic session types (Section~\ref{sec:basic}), we have just two
rules to express subtyping
\begin{mathpar}
  \infer[\m{sub}R]
  {\D \vdash P :: (x : A')}
  {\D \vdash P :: (x : A) & A \leq A'}
  \and
  \infer[\m{sub}L]
  {\D, (x : A) \vdash P :: (z : C)}
  {\D, (x : A') \vdash P :: (z : C) & A \leq A'}
\end{mathpar}
plus the rules for subtyping $A \leq B$.  The latter follow Gay
and Hole~\cite{Gay2005} by defining it coinductively as the largest
type simulation.  This includes the standard notion of depth and width
subtyping for internal and external choice, except that our relation
happens to be exactly reversed from theirs due to our intuitionistic
framework.  We introduce fresh internal definitions for all
intermediate type subexpressions, which allows us a practically
efficient implementation of their algorithm that incrementally
constructs this simulation.  For type-checking purposes, we can
restrict the uses of subtyping to forwarding (rule $\m{id}$), spawn
(rule $\m{def}$), and channel passing (rules ${\tensor}R$ and
${\lolli}L$).  Our implementation of the linear $\lambda$-calculus in
Section~\ref{sec:examples} exploits subtyping by observing that all
values (type \verb`val`) are also expressions (type \verb`exp`), that
is, $\m{val} \leq \m{exp}$.  This means we can pass a channel of type
\verb`val` to a process expecting a channel of type \verb`exp`, which
is used in the implementation of $\beta$-reduction.

The algorithms for subtyping become increasingly more complicated with
the addition of indexed types~\cite{Das20CONCUR} (which is in fact
undecidable) and nested polymorphic types (the subject of ongoing
research, generalizing type equality~\cite{Das21ESOP}).  Nevertheless,
the basic structure of incrementally constructing a simulation remains
intact, just recognizing that a new pair is already in the partial
simulation becomes more difficult.  Our example suite shows that even
in the undecidable cases, type checking (including subtyping) does not
become a significant bottleneck.

\section{Implementation}\label{sec:impl}

We have implemented a prototype for Rast in Standard ML (8100 lines of
code). This implementation contains a lexer and parser (1200 lines),
reconstruction engine (900 lines), an arithmetic solver (1200 lines),
a type checker (2500 lines), pretty printer (400 lines), and an
interpreter (200 lines).  The source code is well-documented and
available open-source~\cite{RastBitBucket}.

\paragraph{\textbf{\textit{Syntax}}}

Table~\ref{tab:syntax} describes the syntax for Rast programs. Each row presents the abstract and
concrete representation of a session type, and its corresponding providing
expression. A program contains a series of mutually recursive type and process
declarations and definitions.
\begin{lstlisting}[caption={Top-Level Declarations},label=env,captionpos=b,belowcaptionskip=-\medskipamount]
  type v[a]...{n}... = A
  decl f[a]...{n}... : (x1 : A1) ... (xn : An) |- (x : A)
  proc x <- f [a]...{n}... x1 ... xn = P
\end{lstlisting}
The first line is a \emph{type definition}, where $v$ is the name with
type parameters $\overline{a}$ and index variables $\overline{n}$ and
$A$ is its definition. The second line is a \emph{process
declaration}, where $f$ is the process name, $(x_1 : A_1) \ldots (x_n : A_n)$ are
the used channels and corresponding types, while the provided channel
is $x$ of type $A$. Finally, the last line is a \emph{process
definition} for the same process $f$ defined using the process
expression $P$. In addition, $f$ can be parameterized by type
variables $\overline{a}$ and index variables $\overline{n}$.  We use a
hand-written lexer and shift-reduce parser to read an input file and
generate the corresponding abstract syntax tree of the program.  The
reason to use a hand-written parser instead of a parser generator is
to anticipate the most common syntax errors that programmers make and
respond with the best possible error messages.

\begin{table}[t]
  \centering
  \begin{tabular}{l l @{\hskip 1em} l l}
  \textbf{Abstract Types} & \textbf{Concrete Types} & \textbf{Abstract Syntax} & \textbf{Concrete Syntax} \\
  \toprule
  $\ichoice{l : A, \ldots}$ & \verb|+{l : A, ...}| & $\esendl{x}{k}$ & \verb|x.k| \\
  $\echoice{l : A, \ldots}$ & \verb|&{l : A, ...}| & $\ecase{x}{\ell}{P}_{\ell \in L}$ & \verb!case x (l => P | ...)! \\
  $A \tensor B$ & \verb|A * B| & $\esendch{x}{w}$ & \verb|send x w| \\
  $A \lolli B$ & \verb|A -o B| & $\erecvch{x}{y}$ & \verb|y <- recv x| \\
  $\one$ & \verb|1| & $\eclose{x}$ & \verb|close x| \\
  & & $\ewait{x}$ & \verb|wait x| \\
  $\texists{a} A$ & \verb|?[a]. A| & $\esendch{x}{[B]}$ & \verb|send x [B]| \\
  $\tforall{a} A$ & \verb|![a]. A| & $\erecvch{x}{[a]}$ & \verb|[a] <- recv x| \\
  $\texists{n} A$ & \verb|?n. A| & $\esendn{x}{e}$ & \verb|send x {e}| \\
  $\tforall{n} A$ & \verb|!n. A| & $\erecvn{x}{n}$ & \verb|{n} <- recv x| \\
  $\tassert{\phi} A$ & \verb|?{phi}. A| & $\eassert{x}{\phi}$ & \verb|assert x {phi}| \\
  $\tassume{\phi} A$ & \verb|!{phi}. A| & $\eassume{x}{\phi}$ & \verb|assume x {phi}| \\
  $\tpaypot{A}{r}$ & \verb!|{r}> A! & $\epay{x}{r}$ & \verb|pay x {r}| \\
  $\tgetpot{A}{r}$ & \verb!<{r}| A! & $\eget{x}{r}$ & \verb|get x {r}| \\
  $\next^t A$ & \verb|({t}) A| & $\edelay{t}$ & \verb|delay {t}| \\
  $\Box A$ & \verb|[] A| & $\ewhen{x}$ & \verb|when x| \\
  $\Dia A$ & \verb|<> A| & $\enow{x}$ & \verb|now x| \\
  $V\indv{A}\indn{e}$ & \verb|V[A]...{e}...| \\
  & & $\fwd{x}{y}$ & \verb|x <-> y| \\
  & & $\procdef{f}{x_1 \ldots x_n}{x}$ & \verb|x <- f x1 ... xn| \\
  \bottomrule
  \end{tabular}
  \caption{Abstract and Corresponding Concrete Syntax for Types and Expressions}%
  \label{tab:syntax}
\end{table}

\paragraph{\textbf{\textit{Validity Checking}}}

Once the program is parsed and its abstract syntax tree is extracted,
we perform a validity check on it. We check that all index
refinements, potentials, and delay operators are non-negative. We also
check that all index expressions are closed with respect to the the
index variables in scope, and similarly for type expressions. To
simplify and improve the efficiency of the subtyping algorithm, we
also assign internal names to type subexpressions~\cite{Das20PPDP,Das21ESOP}
parameterized over their free type and index variables. These internal
names are not visible to the programmer.

\paragraph{\textbf{\textit{Cost Model}}}

The cost model defines the execution cost of each construct. Since our
type system is parametric in the cost model, we allow programmers to
specify the cost model they want to use. Although programmers can
create their own cost model (by inserting $\m{work}$ or $\m{delay}$
expressions in the process expressions), we provide three custom cost
models: $\m{send}$, $\m{recv}$, and $\m{recvsend}$.  If we are
analyzing work (resp.\ time), the $\m{send}$ cost model inserts a
$\m{work \{1\}}$ (resp. $\m{delay} \{1\}$) before (resp.\ after) each
send operation. Similarly, $\m{recv}$ model assigns a cost of $1$ to
each receive operation. The $\m{recvsend}$ cost model assigns a cost
of $1$ to each send and receive operation.

\paragraph{\textbf{\textit{Reconstruction and Type Checking}}}

The programmer can use a flag in the program file to indicate whether
they are using \emph{explicit} or \emph{implicit} syntax. If the
syntax is explicit, the reconstruction engine performs no program
transformation. However, if the syntax is implicit, we use the
implicit type system to approximately type-check the program. Once
completed, we use the forcing calculus, introduced in prior
work~\cite{Das20PPDP} to insert $\m{assert}$,
$\m{assume}$, $\m{pay}$, $\m{get}$ and $\m{work}$ constructs. The core
idea here is simple: insert $\m{assume}$ or $\m{get}$ constructs
eagerly, i.e., as soon as available on a channel, and insert
$\m{assert}$ and $\m{pay}$ lazily, i.e., just before communicating on
that channel.  The forcing calculus proves that this reconstruction
technique is sound and complete in the absence of certain forms of
quantifier alternations (which are checked before reconstruction is
performed).  We only perform reconstruction for proof constraints and
ergometric types; reconstruction of type and index quantifiers and temporal
constructs is left to future work.

The implementation takes some care to provide constructive and precise
error messages, in
particular as session types (not to mention arithmetic refinements,
ergometric types, and temporal types) are likely to be unfamiliar.
One technique is staging: first check approximate type correctness,
ignoring index, ergometric, and temporal types, and only if that check
passes perform reconstruction and strict checking of type.  Another
particularly helpful technique has been \emph{type compression}.  Whenever
the type checker expands a type $V \indv{A} \indn{e}$ with
$V \indv{\alpha} \indn{n} = B$ to
$B[\overline{A}/\overline{\alpha},\overline{e}/\overline{n}]$, we
record a reverse mapping back to $V \indv{A} \indn{e}$. When printing
types for error messages this mapping is consulted, and complex types
may be compressed to much simpler forms, greatly aiding readability of
error messages.  This is feasible in part because all intermediate
subexpressions have an explicit (internal) definition, simplifying the
lookup.
Finally, our implementation uses a bi-directional~\cite{Das20PPDP,Das21ESOP}
type checking algorithm which reconstructs intermediate types for
each channel. This helps localize the source of the error message
as the program point where reconstruction fails. We designed the abstract
syntax tree to also contain the relevant source code location information
which is utilized while generating the error message.

\paragraph{\textbf{\textit{Subtyping}}}

At the core of type checking lies subtyping, defined
coinductively~\cite{Gay2005}. In the presence of arithmetic
refinements, subtyping and also type equality are undecidable, but we
have found what seems to be a practical
approximation~\cite{Das20CONCUR,Das21ESOP}, incrementally
constructing a simulation closed under reflexivity.  The data
structures are rather straightforward, emphasizing simplicity over
efficiency since subtyping tends to be fast in practice.  There are
several places where the translation from the underlying theory to the
implementation are not straightforward.  After the file has been read
and the validity of types has been verified, we compute the
\emph{variance} of all type constructors in all type arguments
(covariant, contravariant, nonvariant, or bivariant) by a greatest
fixed point computation, starting from the assumption that all
arguments are nonvariant.  The variance information is then used when
determining if a new subtyping goal is implied by the partial
simulation constructed so far.  The algorithm employs syntactic
\emph{matching} (allowing the type variables in the simulation to be
instantiated) without consulting the partial simulation again (which
could lead to nontermination).  It also calls upon the constraint
solver to determine if the constrained pairs in the partial
simulation contain enough information to entail the constraints in
the goal.
The theory establishing soundness of this subtyping algorithm in the
presence of nested polymorphic types under their
coinductive interpretation is currently under development.

This algorithm always terminates because we impose a bound on how many
distinct inequalities
$V \indv{A} \indn{e} \leq V' \indv{A'} \indn{e'}$ we consider for any
given pair $V, V'$.  Note that variables also include the internal
names created during elaboration.  However, it may fail to establish
an inequality if the coinductive invariant is not general enough.  Rast
therefore allows the programmer to assert an arbitrary number of
additional type equalities or inequalities with the constructs
\begin{lstlisting}
  eqtype V[A]...{e}... = V'[A']...{e'}...
  eqtype V[A]...{e}... <= V'[A']...{e'}...
\end{lstlisting}
These are abstracted over their free variables, then checked one by
one, assuming all other asserted equalities and inequalities.  The
default construction of the simulation is currently strong enough so
that this feature has not been needed for any of our standard
examples.

\paragraph{\textbf{\textit{Arithmetic Solver}}}

To determine the validity of arithmetic propositions that is used by
our refinement layer, we use a straightforward implementation of
Cooper's decision procedure~\cite{cooper1972theorem} for Presburger
arithmetic.  We found a small number of optimizations were necessary,
but the resulting algorithm has been quite efficient and robust.
\begin{enumerate}
\item We eliminate constraints of the form $x = e$ (where $x$ does not
  occur in $e$) by substituting $e$ for $x$ in all other constraints
  to reduce the total number of variables.

\item We exploit that we are working over natural numbers so all
  solutions have a natural lower bound, i.e., $0$.
\end{enumerate}

We also extend our solver to handle non-linear constraints. Since non-linear
arithmetic is undecidable, in general, we use a normalizer which collects coefficients of
each term in the multinomial expression.
\begin{enumerate}
\item To check $e_1 = e_2$, we normalize $e_1 - e_2$ and check that
  each coefficient of the normal form is $0$.

\item To check $e_1 \geq e_2$, we normalize $e_1 - e_2$ and check that
  each coefficient is non-negative.

\item If we know that $x \geq c$, we substitute $y+c$ for $x$ in the
  constraint that we are checking with the knowledge that the fresh
  $y \geq 0$.

\item We try to find a quick counterexample to validity by plugging
  in $0$ and $1$ for the index variables.
\end{enumerate}
If the constraint does not fall in the above two categories, we print
the constraint and trust that it holds. A user can then view these
constraints manually and confirm their validity. At present, all of
our examples pass without having to trust unsolvable constraints with
our set of heuristics beyond Presburger arithmetic.

\paragraph{\textbf{\textit{Interpreter}}}

The current version of the interpreter pursues a sequential schedule
following a prior proposal~\cite{Pruiksma20arxiv}.  We only execute
programs that have no free type or index variables and only one
externally visible channel, namely the one provided.  When the
computation finishes, the messages that were asynchronously sent along
this distinguished channel are shown, while running processes waiting
for input are displayed simply as a dash '\verb'-''.

The interpreter is surprisingly fast.  For example, using a linear
prime sieve to compute the status (prime or composite) of all numbers
in the range $[2,257]$ takes 27.172 milliseconds using MLton
during our experiments (see machine specifications below).

\section{Further Examples}\label{sec:examples}

We present several different kinds of examples from varying domains
illustrating different features of the type system and algorithms.
Table~\ref{tab:case_study} describes the results: iLOC describes the
lines of source code in implicit syntax, eLOC describes the lines of
code after reconstruction (which inserts implicit constructs), \#Defs
shows the number of process definitions, R (ms) and T (ms) show the
reconstruction and type-checking time in milliseconds respectively.
Note that reconstruction is faster than type-checking since reconstruction
does not involve solving any arithmetic propositions.
The experiments were run on an Intel Core i5 2.7 GHz processor with
16 GB 1867 MHz DDR3 memory.
\begin{enumerate}
  \item \textbf{arithmetic}: natural numbers in unary and binary
  representation indexed by their value and processes implementing
  standard arithmetic operations.

  \item \textbf{integers}: an integer counter represented using two indices
  $x$ and $y$ with value $x-y$.

  \item \textbf{linlam}: expressions in the linear $\lambda$-calculus
  indexed by their size.

  \item \textbf{list}: lists indexed by their size, and some
  standard operations such as \emph{append, reverse, map, fold}, etc.
  Also provides and implementation of stacks and queues using lists.

  \item \textbf{primes}: the sieve of Eratosthenes to
  classify numbers as prime or composite.

  \item \textbf{segments}: type $\m{seg}[n] =
  \forall k. \m{list}[k] \lolli \m{list}[n+k]$ representing partial lists
  with a constant-work append operation.

  \item \textbf{ternary}: natural numbers and integers represented in
  balanced ternary form with digits $0, 1, -1$, indexed by their
  value, and a few standard operations on them.  This example is
  noteworthy since it is the only one stressing the arithmetic
  decision procedure.

  \item \textbf{theorems}: processes representing valid circular~\cite{Derakhshan19corr}
  proofs of simple theorems such as $n(k+1) = nk+n, n+0=n, n*0=0$, etc.

  \item \textbf{tries}: a trie data structure to store multisets of
  binary numbers, with constant amortized work insertion and deletion
  verified with ergometric types.
\end{enumerate}
We highlight interesting examples from
some case studies showcasing the invariants that can be proved using
arithmetic refinements and nested polymorphism.

\begin{table}[t]
  \centering
  \begin{tabular}{l r r r r r}
  \textbf{Module} & \textbf{iLOC} & \textbf{eLOC} & \textbf{\#Defs} & \textbf{R (ms)} & \textbf{T (ms)} \\
  \toprule
  arithmetic & 395 & 619 & 29 & 0.959 & 5.732 \\
  integers & 90 & 125 & 8 & 0.488 & 0.659 \\
  linlam & 88 & 112 & 10 & 0.549 & 1.072 \\
  list & 341 & 642 & 37 & 3.164 & 4.637 \\
  primes & 118 & 164 & 11 & 0.289 & 4.580 \\
  segments & 48 & 76 & 8 & 0.183 & 0.225 \\
  ternary & 270 & 406 & 20 & 0.947 & 140.765 \\
  theorems & 79 & 156 & 13 & 0.182 & 1.095 \\
  tries & 243 & 520 & 13 & 2.122 & 6.408 \\
  \midrule
  \textbf{Total} & \textbf{1672} & \textbf{2820} & \textbf{149} & \textbf{8.883} & \textbf{165.173} \\
  \bottomrule
  \end{tabular}
  \caption{Case Studies}%
  \label{tab:case_study}
\end{table}

\paragraph{\textbf{\textit{Linear \texorpdfstring{$\lambda$}{Lambda}-Calculus}}}

We implemented the linear $\lambda$-calculus with evaluation (weak head normalization) of
terms. We use higher-order abstract syntax, representing
linear abstraction in the object language by a process receiving a
message corresponding to its argument.
This is inspired by
Milner's call-by-name encoding of the (nonlinear) $\lambda$-calculus
in the $\pi$-calculus~\cite{Milner92mscs}.  We expand on it by considering
typing in metalanguage, and also provide a static analysis of size of
normal forms and number of reductions.
\begin{lstlisting}
type exp = +{ lam : exp -o exp,
                app : exp * exp }
\end{lstlisting}
We would like evaluation to return a value (a $\lambda$-abstraction),
so we take advantage of the structural nature of types (allowing
us to reuse the label $\mb{lam}$) to define
the value type.
\begin{lstlisting}
type val = +{ lam : exp -o exp }
\end{lstlisting}
Rast can infer that $\m{val}$ is a subtype of $\m{exp}$.
We can derive
constructors $\mi{apply}$ for expressions and $\mi{lambda}$
for values (we do not need the corresponding constructor for
expressions).
\begin{lstlisting}
decl apply : (e1 : exp) (e2 : exp) |- (e : exp)
proc e <- apply e1 e2 =
  e.app ; send e e1 ; e <-> e2

decl lambda : (f : exp -o exp) |- (v : val)
proc v <- lambda f = v.lam ; v <-> f
\end{lstlisting}
As a simple example, we follow the representation
of a combinator that swaps the arguments to a function.
\begin{lstlisting}
(* swap = \f. \x. \y. (f y) x *)
decl swap : . |- (e : exp)
proc e <- swap =
  e.lam ; f <- recv e ;
  e.lam ; x <- recv e ;
  e.lam ; y <- recv e ;
  fy <- apply f y ;
  e <- apply fy x
\end{lstlisting}
Evaluation is now the following very simple process.
\begin{lstlisting}
decl eval : (e : exp) |- (v : val)
proc v <- eval e =
  case e ( lam => v <- lambda e
          | app => e1 <- recv e ;     % e = e2
                    v1 <- eval e1 ;
                    case v1 ( lam => send v1 e ;
                                       v <- eval v1 ) )
\end{lstlisting}
If $e$ sends a $\mb{lam}$ label, we just rebuild the expression as a
value.  If $e$ sends an $\mb{app}$ label then $e$ represents a linear
application $e_1\, e_2$ and the continuation has type
$\m{exp} \tensor \m{exp}$.  This means we \emph{receive} a channel
representing $e_1$ and the continuation (still called $e$) behaves
like $e_2$.  We note this with a comment in the source.  We then
evaluate $e_1$ which exposes a $\lambda$-expression along the channel
$v_1$.  We send $e$ along $v_1$, carrying out the reduction via
communication.  The result of this (still called $v_1$) is evaluated
to yield the final value $v$.
This program is available in the repository at \verb'examples/linlam.rast'.

We would now like to prove that the value of a linear
$\lambda$-expression is smaller than or equal to the original
expression.  At the same time we would like to rule out a class of
so-called \emph{exotic terms} in the representation, which are
possible due to the presence of recursion in the metalanguage.  We
achieve this by indexing the types $\m{exp}$ and $\m{val}$ with their
\emph{size}.  For an application, this is easy: the size is one more
than the sum of the sizes of the subterms.
\begin{lstlisting}
type exp{n} = +{ lam : ...
                   app : ?n1. ?n2. ?{n = n1+n2+1}. exp{n1} * exp{n2} }
\end{lstlisting}
The size $n_2+1$ of a $\lambda$-expression is one more than the size $n_2$ of
its body, but what is that in our higher-order representation? The
body of a linear function takes an expression of size $n_1$ and then
behaves like an expression of size $n_1 + n_2$.  Solving for $n_2$
then gives use the following type definitions and types for
the constructor processes.
\begin{lstlisting}
type exp{n} = +{lam : ?{n > 0}. !n1.exp{n1} -o exp{n1+n-1},
                  app : ?n1. ?n2. ?{n = n1+n2+1}. exp{n1} * exp{n2}}

type val{n} = +{ lam : ?{n > 0}. !n1.exp{n1} -o exp{n1+n-1} }

decl apply{n1}{n2} : (e1 : exp{n1}) (e2 : exp{n2}) |- (e : exp{n1+n2+1})
decl lambda{n2} : (f : !n1. exp{n1} -o exp{n1+n2}) |- (v : val{n2+1})
\end{lstlisting}
The universal quantification over $n_1$ in the type of $\mb{lam}$ is
important, because a linear $\lambda$-expression may be applied to an
argument of any size. We also cannot predict the size of the result of
evaluation, so we have to use existential quantification: The value of
an expression of size $n$ will have size $k$ for some $k \leq n$.
\begin{lstlisting}
decl eval{n} : (e : exp{n}) |- (v : ?k. ?{k <= n}. val{k})
\end{lstlisting}
Because witnesses for quantifiers are not reconstructed, the
evaluation process has to send and receive suitable sizes.
\begin{lstlisting}
proc v <- eval{n} e =
  case e ( lam => send v {n} ;
                  v <- lambda{n-1} e
          | app => {n1} <- recv e ;
                    {n2} <- recv e ;
                    e1 <- recv e ;
                    v1 <- eval{n1} e1 ;
                    {k2} <- recv v1 ;
                    case v1 ( lam => send v1 {n2} ;
                                       send v1 e ;
                                       v2 <- eval{n2+k2-1} v1 ;
                                       {l} <- recv v2 ;
                                       send v {l} ; v <-> v2))
\end{lstlisting}
Type-checking now verifies that if evaluation terminates, the
resulting value is smaller than the expression (or of equal size).
The repository contains the implementation in the file
\verb'examples/linlam-size.rast'.

Remarkably, ergometric session types can bound the number of reductions using an amortized analysis
of work! For this, we assign 1 erg (unit of potential) to each
$\lambda$-expression. Our cost model is that all operations are free,
except the equivalent of a $\beta$-reduction which costs 1 erg.
Because transfer of potential is reconstructed, the program is
very close to the original, size-free program.
\begin{lstlisting}
type exp = +{ lam : |> exp -o exp,
                app : exp * exp }

type val = +{ lam : |> exp -o exp }
\end{lstlisting}
The \verb'|>' in the $\mb{lam}$ branch denotes a unit potential
associated with an $\lambda$-expression.
The definitions for $\mi{apply}$ and $\mi{lambda}$ are unchanged.
\begin{lstlisting}
decl apply : (e1 : exp) (e2 : exp) |- (e : exp)
proc e <- apply e1 e2 =
  e.app ; send e e1 ; e <-> e2

decl lambda : (f : exp -o exp) |{1}- (v : val)
proc v <- lambda f =
  v.lam ; v <-> f
\end{lstlisting}
The \verb'|{1}-' denotes a unit potential on the $\mi{lambda}$
process. This potential is consumed to send 1 erg after sending
the $\mb{lam}$ label. Note that reconstruction enables the programmer
to skip such work constructs.
\begin{lstlisting}
decl eval : (e : exp) |- (v : val)
proc v <- eval e =
  case e ( lam => v <- lambda e
         | app => e1 <- recv e ;      % e = e2
                   v1 <- eval e1 ;
                   case v1 ( lam => work ;
                                      send v1 e ; % beta
                                      v <- eval v1 ) )
\end{lstlisting}
Type-checking here verifies that the reduction of a given expression
with $n$ $\lambda$-abstractions to a value performs at most $k < n$
$\beta$-reductions, with a potential of $n-k$ for further reductions
remaining in the value. This means that there are exactly $n-k$
$\lambda$-abstractions remaining in the result.
The only change in the program is the $\m{work}$ construct to realize
the cost model of only counting the reductions.
Rast allows programmers to choose from one of the existing cost models
or provide their own custom cost model.
Again, the remaining constructs (e.g.\ paying and getting potential)
are automatically inserted by the reconstruction engine.
The full code can be found in the file \verb'examples/linlam-reds.rast'
in the repository.

\paragraph{\textbf{\textit{Trie Data Structure}}}

We now implement multisets of natural numbers (in binary form), as
introduced at the end of Section~\ref{sec:refine}.  One of the key
questions is how to maintain linearity in the design of the data
structure and interface.  For example, should we be able to delete an
element from the trie, not knowing a priori if it is even \emph{in}
the trie?  To avoid exceedingly complex types to account for these
situations, the process maintaining a trie offers an interface with
two operations: insert (label $\mb{ins}$) and delete (label
$\mb{del}$).  We index the type \verb'trie{n}' with the number of
elements in the trie, so inserting an element always increases $n$ by
1.  If the element is already present, we just add 1 to its
multiplicity.  Deleting an element actually removes all copies of it
and returns its multiplicity $m$.  If the element is \emph{not} in the
trie, we just return a multiplicity of $m = 0$.  In either case, the
trie contains $n-m$ elements afterwards.
\begin{lstlisting}
type trie{n} = &{ins : !k. bin{k} -o trie{n+1},
                   del : !k. bin{k} -o ?m. ?{m <= n}. bin{m} * trie{n-m}}
\end{lstlisting}
This type requires universal quantification over $k$, (written \verb'!k') which is the
value of the number inserted into or deleted from the trie on each
interaction (which is arbitrary).

The basic idea
of the implementation is that each bit in the number
\verb'x : bin{k}' addresses a subtrie: if it is $\mb{b0}$ we descend into
the left subtrie, if it is $\mb{b1}$ we descent into the right
subtrie.  If it is $\mb{e}$ we have found (or constructed) the node
corresponding to $x$ and we either increase its multiplicity (for
insert), or respond with its multiplicity and set the new multiplicity
to zero (for delete).  We have two forms of processes: a \emph{leaf}
with zero elements and an interior node with $n_0+m+n_1$ elements
(where $n_0$ and $n_1$ and the number of elements in the left and
right subtries, and $m$ is the multiplicity of the number
corresponding to this node in the trie).
\begin{lstlisting}
decl leaf : . |- (t : trie{0})
decl node{n0}{m}{n1} :
  (l : trie{n0}) (c : ctr{m}) (r : trie{n1}) |- (t : trie{n0+m+n1})
\end{lstlisting}
The code is somewhat repetitive, so we only show the code
for inserting an element into an interior node.
\begin{lstlisting}
proc t <- node{n0}{m}{n1} l c r =
  case t (
    ins => {k} <- recv t ;
           x <- recv t ;
           case x ( b0 => {k'} <- recv x ;
                          l.ins ; send l {k'} ; send l x ;
                          t <- node{n0+1}{m}{n1} l c r
                  | b1 => {k'} <- recv x ;
                          r.ins ; send r {k'} ; send r x ;
                          t <- node{n0}{m}{n1+1} l c r
                  | e => wait x ;
                         c.inc ;
                         t <- node{n0}{m+1}{n1} l c r )
  | del => ...)
\end{lstlisting}
What does type-checking verify in this case?  It shows that the number
of elements in the trie increases and decreases as expected for each
insert and delete operation.  On the other hand, it does not verify
that the \emph{correct} multiplicities are incremented or decremented,
which is beyond the reach of the current type system.
The source code is available at \verb'examples/trie-work.rast'.

\paragraph{\textbf{\textit{Expression Server}}}
Nested polymorphism in Rast can also be employed to
ensure interesting invariants.
As an example, we adapt the example of an arithmetic expression
from prior work on context-free session types~\cite{Thiemann16icfp}.
The type of the server is defined as
\begin{lstlisting}
  type bin = +{ b0 : bin, b1 : bin, e : 1 }
  type tm[K] = +{ const : bin * K,
                    add : tm[tm[K]],
                    double : tm[K] }
\end{lstlisting}
The type \verb`bin` represents a binary number as before, except we
dropped the index in order to focus on nested polymorphism.

An arithmetic term, parameterized by continuation type \verb`K` can
have one of three forms: a constant, the sum of two terms,
or the double of a term. Consequently, the type \verb`tm[K]` ensures that
 a process providing \verb`tm[K]` is a \emph{well-formed term}: it
either sends the \verb`const` label followed by sending a constant binary
number of type \verb`bin` and continues with type \verb`K`; or it sends
the \verb`add` label and continues with \verb`tm[tm[K]]`, where
the two terms denote the two summands; or it sends the \verb`double`
label and continues with \verb`tm[K]`.
 In particular, the continuation type
\verb`tm[tm[K]]` in the \verb`add` branch enforces that the process must send exactly
two summands for sums.

As a first illustration, consider two binary constants $a$ and $b$,
and suppose that we want to create the expression $a+2b$. We can issue
commands to the expression server in a \emph{prefix notation}
to obtain $a+2b$, as shown in the following \verb`exp[K]` process, which is
parameterized by a continuation type \verb`K`.
\begin{lstlisting}
  decl exp[K] : (a : bin) (b : bin) (k : K) |- (e : tm[K])
  proc e <- exp[K] a b k =
    e.add ;                 % (a:bin) (b:bin) (k:K) |- (e : tm[tm[K]])
    e.const ; send e a ;    % (b:bin) (k:K) |- (e : tm[K])
    e.double ;              % (b:bin) (k:K) |- (e : tm[K])
    e.const ; send e b ;    % (k:K) |- (e : K)
    e <-> k
\end{lstlisting}
In prefix notation, $a+2b$ would be written $+ \; (a) \; (2 \; b)$,
which is exactly the form followed by the \verb`exp` process:
The process sends \verb`add`, followed by \verb`const` and the number
\verb`a`, followed by \verb`double`, \verb`const`, and \verb`b`.
Finally, the process continues at type \verb`K` by forwarding \verb`k` to \verb`e`.
To assist the reader, we describe the intermediate typing contexts on the
right that are automatically reconstructed by the Rast type checker.

The type \verb`tm[K]` ensures that any process offering it must
be a \emph{well-formed term}. In particular, the continuation type
\verb`tm[tm[K]]` in the \verb`add` branch enforces that once
the \verb`add` command is issued, the process must send exactly
two summands. Similar properties hold of the \verb`const`
and \verb`double` branches.

To evaluate a term, we can define an \verb`eval` process, parameterized by type \verb`K`:

\begin{lstlisting}
decl eval[K] : (t : tm[K]) |- (v : bin * K)
\end{lstlisting}
The \verb`eval` process uses channel
\verb`t : tm[K]` as argument,
and offers \verb`v : bin * K`. The process evaluates
term \verb`t` and sends its binary value along \verb`v`.
\begin{lstlisting}
  decl eval[K] : (t : tm[K]) |- (v : bin * K)
  proc v <- eval[K] t =
    case t (
      const =>                  % (t : bin * K) |- (v : bin * K)
        n <- recv t ;           % (n : bin) (t : K) |- (v : bin * K)
        send v n ; v <-> t
    | add =>                    % (t : tm[tm[K]]) |- (v : bin * K)
        v1 <- eval[tm[K]] t ;   % (v1 : bin * tm[K]) |- (v : bin * K)
        n1 <- recv v1 ;         % (n1 : bin) (v1 : tm[K]) |- (v : bin * K)
        v2 <- eval[K] v1 ;      % (n1 : bin) (v2 : bin * K) |- (v : bin * K)
        n2 <- recv v2 ;         % (n1 : bin)(n2 : bin)(v2 : K) |- (v : bin * K)
        n <- plus n1 n2 ;       % (n : bin) (v2 : K) |- (v : bin * K)
        send v n ; v <-> v2
    | double =>                 % (t : tm[K]) |- (v : bin * K)
        v1 <- eval[K] t ;       % (v1 : bin * K) |- (v : bin * K)
        n1 <- recv v1 ;         % (n1 : bin) (v1 : K) |- (v : bin * K)
        n <- double n1 ;        % (n : bin) (v1 : K) |- (v : bin * K)
        send v n ; v <-> v1
    )
\end{lstlisting}
Intuitively, the process evaluates
term \verb`t` and sends its binary value along \verb`v`. If \verb`t` is a constant, then
\verb`eval[K]` receives the constant $n$, sends it along $v$ and forwards.

The interesting case is the \verb`add` branch. We evaluate the first
summand by spawning a new \verb`eval[K]` process on \verb`t`. Note that since
the type of \verb`t` (indicated on the right) is \verb`tm[tm[K]]`
and hence, the recursive call to \verb`eval` is at parameter  \verb`tm[K]`.
We store the value of the first summand at channel \verb`n1 : bin`.
Then, we continue to evaluate the second summand by calling \verb`eval[K]`
on \verb`t` again and storing its value in \verb`n2 : bin`. Finally, we add
\verb`n1` and \verb`n2` by calling the \verb`plus` process (which has
a straightforward definition), and send the result
\verb`bin` along \verb`v`. We follow a similar approach for the \verb`double`
branch.

\section{Related Work}\label{sec:related}

The literature on session types is by now vast, so we focus our review
of related work on \emph{binary session types} (rather than multiparty
session types) with \emph{implementations} (rather than theoretical
foundations).  Among them, we can distinguish those that offer a
library or embedding to a pre-existing language, and those that may be
considered stand-alone language designs.

\paragraph{\textbf{\textit{Libraries}}}
There are a number of libraries for session types.
Such libraries tend to have a very different flavor from Rast
because they focus on practical usability in the context of a
general-purpose language.  As such, the challenge usually is how to
encode session types so programs can be statically checked against
them and how to achieve the expected dynamic behavior.  Among them we
find libraries for Haskell~\cite{Lindley16haskell,Orchard16popl},
Scala~\cite{Scalas16ecoop}, OCaml~\cite{Padovani17jfp}, and
Rust~\cite{Jespersen15wgp}.  Noteworthy is the embedding of session
types in ATS~\cite{Xi16arxiv} because, unlike the others, ATS supports
arithmetic indexing similar to Rast.  The most recent library for
Rust~\cite{Chen20arxiv} is perhaps the closest to Rast in that it
extends the exact basic system of session types from
Section~\ref{sec:basic} with shared types~\cite{Balzer17icfp}.  While
some of these libraries permit limited polymorphism, none of them support
ergometric or temporal types.

\paragraph{\textbf{\textit{Languages}}}
Designing complete languages like Rast frees
the researcher from the limitations and idiosyncrasies of the host
language as they explore the design space.  A relatively early effort
was the object-oriented language MOOL~\cite{Vasconcelos11beatcs} which
distinguishes linear and nonlinear channels.

A different style of language is
SePi~\cite{Baltazar12linearity,Franco13sefm} based on the
$\pi$-calculus.  It supports \emph{linear} refinements in terms of
uninterpreted propositions (which may reference integers) in addition
to assert and assume primitives on them.  They are not intended to
capture internal properties of data structures of processes; instead,
they allow the programmer to express some security properties.

The CO$_2$ middleware language~\cite{Bartoletti15forte,Bartoletti15facs}
supports \emph{binary timed session types}.  The notion of time here
is \emph{external}.  As such, it does not measure work or span based
on a cost model like Rast, but specifies interaction time windows for
processes that can be enforced dynamically via monitors.

Concurrent C0~\cite{Willsey16linearity} is an implementation of linear
and shared session types as an extension of C0, a small type-safe and
memory-safe subset of C\@.  It integrates the basic session types from
Section~\ref{sec:basic} with shared session type~\cite{Balzer17icfp}
in the context of an imperative language.
Relatedly, the Nomos language~\cite{Das21CSF} integrates linear and
shared ergometric session types with a functional language
to aid smart contract programming.
Although Nomos does not support temporal types and polymorphism,
it embeds a linear programming solver to automatically infer the
exact potential annotations.

Links~\cite{Lindley16icfp,Lindley17book,Fowler19popl} is a language
aimed at developing web applications.  While based on a different
foundations, it is related to SILL~\cite{Toninho13esop,Griffith16phd}
in that both integrate traditional functional types with linear
session types.  As such, they can express many (nonlinear) programs
that Rast cannot, but they support neither arithmetic refinements nor
ergometric or temporal types.

Context-free session types~\cite{Thiemann16icfp,Almeida20tacas}
generalize ordinary session types with sequential composition as well
as permitting some polymorphism.  The linear sublanguage of
context-free session types can be modeled in Rast with nested
polymorphism~\cite{Das21ESOP}.
Several dependent extensions for session types in prior work
including proof exchanges~\cite{ToninhoPPDP11},
subtyping and constraint relations based on ATS~\cite{WuX17DST},
and equality based on $\beta\eta$-congruences~\cite{ToninhoFOSSACS18}.
However, none of them formally investigate type equality, nor provide
a type checking algorithm realized in an implementation.

\section{Conclusion}\label{sec:conc}

This paper describes the Rast programming language. In particular, we
focused on the concrete syntax, type checking and subtyping,
parametric polymorphism~\cite{Das21ESOP}, the
refinement layer~\cite{Das20CONCUR,Das20PPDP}, and its applicability to
work~\cite{Das18lics} and span analysis~\cite{Das18icfp}. The
refinements rely on an arithmetic solver based on Cooper's
algorithm~\cite{cooper1972theorem}. The
interpreter uses the shared memory semantics introduced in recent
work~\cite{Pruiksma20arxiv}. We concluded with several examples
demonstrating the efficacy of the refined type system in expressing
and verifying properties about data structure sizes and values.
We also illustrated the work and span bounds for several examples,
all of which have been verified with our system, and are available
in an open-source repository~\cite{RastBitBucket}.

In the future, we plan to address some limitations of the Rast language.
One goal of Rast was to explore the boundaries of purely linear
programming with general recursion.
Often, this imposes a certain programming discipline and can be
inconvenient if we need to drop or duplicate channels.
Recent work on adjoint logic~\cite{Pruiksma19places} uniformly
integrates different logical layers into a unified language by
assigning modes to communication.
We plan to utilize this adjoint formulation to support shared~\cite{Balzer17icfp}
and unrestricted channels.
Prior work on the SILL~\cite{Griffith16phd} and Nomos~\cite{Das21CSF}
have demonstrated such an integration is helpful in general-purpose
programming.

In the direction of parametric polymorphism, we plan to develop the
theory of subtyping. Our initial investigation suggests that subtyping
is undecidable, and thus would like to explore the boundaries of our
current subtyping algorithm. Relatedly, we also plan to explore if
we can extend the ideas of reconstruction to explicit quantifiers.

With respect to refinements, we intend to pursue richer
constraint domains such as non-linear arithmetic, particularly SMT\@.
We would also like to support reconstruction for the temporal fragment
of Rast.
Unfortunately, the $\Next$ operator affects all connected channels
at once and its proof-theoretic properties are not as uniform as those
of polymorphism, proof constraints, or ergometric types, posing a significant challenge.
We also plan to explore dependent session type systems to express
data-dependent distributed protocols.

\bibliographystyle{alphaurl}
\bibliography{refs}

\end{document}